\documentclass{jfm}

\usepackage{subcaption}

\usepackage{graphicx}

\usepackage{natbib,amsmath}
\usepackage{mathtools,bbm}
\usepackage[dvipsnames]{xcolor}
\usepackage{hyperref}
\usepackage[utf8]{inputenc}

\definecolor{light-gray}{gray}{0.95}
\ifCUPmtlplainloaded \else
  \checkfont{eurm10}
  \iffontfound
    \IfFileExists{upmath.sty}
      {\typeout{^^JFound AMS Euler Roman fonts on the system,
                   using the 'upmath' package.^^J}%
       \usepackage{upmath}}
      {\typeout{^^JFound AMS Euler Roman fonts on the system, but you
                   dont seem to have the}%
       \typeout{'upmath' package installed. JFM.cls can take advantage
                 of these fonts,^^Jif you use 'upmath' package.^^J}%
      }
  \else
  \fi
\fi

\ifCUPmtlplainloaded \else
  \checkfont{msam10}
  \iffontfound
    \IfFileExists{amssymb.sty}
      {\typeout{^^JFound AMS Symbol fonts on the system, using the
                'amssymb' package.^^J}%
       \usepackage{amssymb}%
         \let\leq=\leqslant
         \let\geq=\geqslant
      }{}
  \fi
\fi

\ifCUPmtlplainloaded \else
  \IfFileExists{amsbsy.sty}
    {\typeout{^^JFound the 'amsbsy' package on the system, using it.^^J}%
     \usepackage{amsbsy}}
    {\providecommand\boldsymbol[1]{\mbox{\boldmath $##1$}}}
\fi

\newcommand\solidrule[1][1cm]{\rule[0.5ex]{#1}{.4pt}}
\newcommand\dashedrule{\mbox{%
  \solidrule[1mm]\hspace{1mm}\solidrule[1mm]\hspace{1mm}\solidrule[1mm]}}
  \newcommand\dottedrule{\mbox{%
  $\cdot$\hspace{1mm}$\cdot$\hspace{1mm}$\cdot$}}
    \newcommand\dasheddottedrule{\mbox{%
  \solidrule[1mm]\hspace{1mm}$\cdot$\hspace{1mm}\solidrule[1mm]}}

\newsavebox{\astrutbox}
\sbox{\astrutbox}{\rule[-5pt]{0pt}{20pt}}

  \newcommand{\rmd}{{\rm d}}
\newcommand{\bx}{{ \boldsymbol{x} }}

\newcommand{\br}{{ \boldsymbol{r} }}

\newcommand{\by}{{ \boldsymbol{y} }}

\newcommand{\RR}{{\mathbb{R}}}

\newcommand{\bW}{{\wt{\mathbf{W}}}}
\newcommand{\bu}{{ \boldsymbol{u}}}
\newcommand{\bw}{{ \boldsymbol{w}}}
\newcommand{\bxi}{{\mbox{\boldmath $\xi$}}}
\newcommand{\bpi}{{\mbox{\boldmath $\pi$}}}
\newcommand{\bomega}{{\mbox{\boldmath $\omega$}}}
\newcommand{\brho}{{\mbox{\boldmath $\rho$}}}

\newcommand{\hd}{\hat{\rmd}}
\newcommand{\tbxi}{\tilde{\bxi}}
\newcommand{\tbpi}{\tilde{\bpi}}

\newcommand{\bE}{{\mathbb{E}}}

\newcommand{\wt}{\widetilde}

\newcommand{\tth}{{\tilde{\theta}}}

\title[Fluctuation-Dissipation Relation]{A Lagrangian fluctuation-dissipation relation for scalar turbulence, I. \\
       Flows with no bounding walls.}

\author[T. D. Drivas and G. L. Eyink]%
{Theodore D. Drivas$^1$ and Gregory L. Eyink $^{1,2}$}

\affiliation{$^1$Department of Applied Mathematics \& Statistics, The Johns Hopkins University, Baltimore, MD 21218, USA\\[\affilskip]
$^2$Department of Physics \& Astronomy, The Johns Hopkins University, Baltimore, MD 21218, USA}

\pubyear{2010}
\volume{650}
\pagerange{119--126}
\date{?; revised ?; accepted ?. - To be entered by editorial office}
\begin{document}

\maketitle

\begin{abstract}
{An exact relation is derived between scalar dissipation due to molecular diffusivity and 
the randomness of stochastic Lagrangian trajectories for flows without bounding walls. This ``Lagrangian fluctuation-dissipation relation'' 
equates the scalar dissipation for either passive or active scalars to the variance of scalar inputs associated to initial scalar 
values and internal scalar sources, as those are sampled backward in time by the stochastic Lagrangian trajectories. 
As an important application, we reconsider the phenomenon of  ``Lagrangian spontaneous stochasticity''
or persistent non-determinism of Lagrangian particle trajectories in the limit of vanishing viscosity and diffusivity. 
Previous work on the Kraichnan (1968) model of turbulent scalar advection has shown that anomalous scalar dissipation
is associated in that model to Lagrangian spontaneous stochasticity. There has been controversy, however, 
regarding the validity of this mechanism for scalars advected by an actual turbulent flow. We here completely resolve
this controversy by exploiting the fluctuation-dissipation relation. For either a passive or active scalar
advected by any divergence-free velocity field, including solutions of the incompressible Navier-Stokes equation, and 
away from walls, we prove that anomalous scalar dissipation requires Lagrangian spontaneous stochasticity.  
For passive scalars we prove furthermore that spontaneous stochasticity yields anomalous dissipation for 
suitable initial scalar fields, so that the two phenomena are there completely equivalent. These points 
are illustrated by numerical results from a database of homogeneous, isotropic turbulence, which provide both 
additional support to the results and physical insight into the representation of diffusive effects by stochastic 
Lagrangian particle trajectories.} 
\end{abstract}



\section{Introduction }

A fundamental feature of turbulent flows is the enhanced dissipation of kinetic energy. It was suggested by 
G. I. \cite{taylor1917observations} that kinetic energy can ``be dissipated in fluid of infinitesimal viscosity''.
 This idea that turbulent dissipation might become independent of molecular viscosity at sufficiently
high Reynolds numbers was pursued further by  \cite{kolmogorov1941dissipation,kolmogorov1941local,kolmogorov1941equations}
and \cite{onsager1945distribution,Onsager49} in developing their theories of turbulence. The physical phenomenon 
is sometimes called the turbulent ``dissipative anomaly'' or even the ``zeroth law of turbulence'', although it 
is, of course, no ``law'' but rather an experimentally observed phenomenon which is still only partially understood theoretically. 
For current empirical evidence from laboratory experiments and numerical simulations, see e.g. 
\cite{Kanedaetal03,Pearsonetal02,sreenivasan1998update}, whose data are all consistent with energy dissipation in the bulk of turbulent flows being essentially 
independent of viscosity. Similar phenomena are expected for other turbulent systems, in particular for scalar fields advected 
by a turbulent fluid, such as concentrations of dyes or aerosols, temperature fluctuations, etc. It was suggested
by \cite{Taylor1922diffusion} that diffusion by turbulence should depend ``little on the molecular conductivity and viscosity 
of the fluid'' and the asymptotic independence of the dissipation rate of scalar fluctuations from 
the molecular transport coefficients was a fundamental assumption in the Kolmogorov-style theories of scalar 
turbulence developed by \cite{Obukhov49} and \cite{corrsin1951spectrum}. A very comprehensive review of the empirical 
evidence for this hypothesis on scalar dissipation is contained in the paper of \cite{donzis2005scalar}, whose 
compilation of data is again consistent with scalar dissipation in the bulk of turbulent flows being insensibly 
dependent on molecular transport coefficients at sufficiently high Reynolds and P\'eclet numbers. This 
phenomenon still requires a complete theoretical explanation.  
  
Fundamental new ideas on the Lagrangian origin of turbulent scalar dissipation arose from mathematical 
work of \cite{Bernardetal98}, which was carried out in the \cite{Kraichnan68} model of synthetic  turbulence. 
In this model the advecting velocity is a Gaussian random field that has Kolmogorov-type scaling of increments in 
space but a white-noise correlation in time. It was shown in the Kraichnan model that the dissipative anomaly 
for a decaying passive scalar is due to a remarkable phenomenon called {\it spontaneous stochasticity}
\citep{chaves2003lagrangian}. Simply stated, \cite{Bernardetal98} showed that Lagrangian particle trajectories 
become non-unique and stochastic in the infinite Reynolds-number limit for a {\it fixed} initial particle 
position and a {\it fixed} velocity realization, due to the spatial roughness of the advecting velocity field.
More precisely, they showed that at very large Reynolds and P\'eclet numbers, when the velocity 
field is smooth but approximates a ``rough'' field over a long range of scales, small stochastic perturbations 
on Lagrangian trajectories due to molecular diffusivity lead to persistent randomness over any finite times
even as the perturbations vanish. This effect is due to the explosive (super-ballistic) dispersion of particle pairs
in a turbulent flow predicted by \cite{Richardson26}, which leads to loss of memory of initial particle 
separations or of amplitudes of stochastic perturbations. For excellent reviews of this and related studies on the 
Kraichnan model, see \cite{Falkovichetal01, Kupiainen03, gawedzki08}. 

Since this pioneering work, however, there have been recurrent 
doubts expressed concerning the validity of these results for real hydrodynamic turbulence. For example,  
\cite{tsinober2009informal} (section 5.4.5) has argued that in real fluids ``the flow field is smooth. In such flows 
`phenomena' like `spontaneous stochasticity' and `breakdown of Lagrangian flow' do not arise and one has 
to look at different more realistic possibilities.'' This is a simple misunderstanding, because 
spontaneous stochasticity is a phenomenon that appears for smooth velocity fields that merely appear 
``rough'' over a long range of scales. More serious questions have been raised concerning the approximation 
of a white-noise temporal correlation in the Kraichnan model.  In a recent detailed comparison of passive 
scalars in the Kraichnan model and in fluid turbulence, \cite{sreenivasan2010lagrangian} have remarked that 
``It is still unclear in the Kraichnan model as to which 
 qualitative and quantitative differences arise from the finite-time correlation of the advecting flow.''  This 
 latter paper discussed also some of the challenges in extending results for the Kraichnan model to 
 understanding of the energy cascade in Navier-Stokes turbulence. 
 
{The principal contribution of the present paper is a new approach to the theory of turbulent scalar dissipation 
based upon an exact {\it fluctuation-dissipation relation for scalars}. Our new relation expresses an
equality between the time-averaged scalar dissipation and the input of scalar variance from 
the initial data and interior scalar sources, as these are sampled by stochastic Lagrangian 
trajectories. This relation makes it intuitively clear that scalar dissipation requires non-vanishing 
Lagrangian stochasticity. In fact, using our new relation, we can prove rigorously the following fact: 
{\it Away from boundaries and for any advecting velocity field whatsoever, spontaneous stochasticity 
of Lagrangian particle trajectories is sufficient for anomalous dissipation of passive scalars, and necessary 
for anomalous dissipation of both passive and active scalars.} Thus, there is no possible mechanism for 
a scalar dissipative anomaly in such situations other than spontaneous stochasticity. In this way we completely 
resolve the controversies on the applicability of the dissipation mechanisms in the Kraichnan model to 
scalars in hydrodynamic turbulence, at least away from walls. 
The importance of our exact fluctuation-dissipation relation (FDR) is not limited to analysis 
of anomalous scalar dissipation and it is valid even when scalar dissipation may vanish as $\nu,\kappa\rightarrow 0$. 
In general our relation gives a new Lagrangian viewpoint on dissipation of scalars, both active and passive. As such, it 
generalizes some previously derived relations, such as that of \cite{Sawfordetal05, Sawfordetal16} for scalars forced 
by a mean scalar gradient and the exact balance relations for stochastic scalar sources which are Gaussian 
white-in-time \citep{novikov1965functionals}. In two companion papers (\cite{PaperII,PaperIII}; hereafter denoted II, III), we show how 
the FDR extends also to wall-bounded domains, with either fixed-scalar (Dirichlet) or fixed-flux (Neumann) conditions 
for the scalar field, and we apply the FDR to the concrete problem of Nusselt-Rayleigh scaling in turbulent Rayleigh-B\'enard convection.}

{The detailed contents of the present paper are as follows: In section \ref{sect:ss1} we first derive the stochastic representation 
of scalar advection and our FDR, in case of flows in domains without walls. In section \ref{BGK} we review the notion of 
spontaneous stochasticity, with numerical verifications from a database of homogeneous, isotropic turbulence.
In section \ref{sec:SS-AD} we establish the connection of spontaneous stochasticity with anomalous scalar dissipation. 
In the summary and discussion section \ref{sec:summary} we discuss both the implications for turbulent vortex dynamics 
and other Lagrangian aspects of turbulence, and also the outstanding challenges, including that of relating spontaneous stochasticity 
to anomalous dissipation of kinetic energy. Three appendices give further details, including rigorous mathematical 
proofs of all of the results in the main text. These deal with the connection between spontaneous stochasticity and 
anomalous dissipation (Appendix \ref{proofs}),
the relation of our scalar FDR to previous results in the literature (Appendix \ref{ensembles}),
and discussion of numerical methods employed (Appendix \ref{numerics}). Footnotes are scattered 
throughout the text, which provide important details for specialists. The general reader can ignore most of these 
in a first passage through the paper and still gain an overall understanding of the contents.} 



\section{Lagrangian Fluctuation Dissipation Relation}\label{sect:ss1}

We consider in this paper turbulent fluid flows in finite domains without walls.  A relevant numerical 
example is DNS of turbulence in a periodic box.  A set of examples from Nature is provided by large-scale flows 
in thin planetary atmospheres, which can be modeled as 2D flows on a sphere. Mathematically speaking, 
the results in this section apply to fluid flows on any compact Riemannian manifold without boundary, merely replacing 
the Wiener process with the Brownian motion on the manifold whose infinitesimal generator is the Laplace-Beltrami 
operator \citep{IkedaWatanabe89}. For simplicity of presentation, we derive the relation only for periodic domains.   

Scalar fields $\theta$ (such as temperature, dye or pollutants) transported by a fluid with velocity $\bu$ are described 
by the advection-diffusion equation 
\begin{align}\label{eq:PS}
\partial_t \theta+\bu\cdot \nabla \theta &= \kappa \Delta \theta + S
\end{align}
with $S(\bx,t)$ a source field and with $\kappa>0$ the molecular diffusivity of the scalar.  The work of \cite{Bernardetal98} 
employed a stochastic representation of the solutions of this equation which is known as the Feynman-Kac representation 
in the mathematics literature \citep{oksendal2013stochastic} and as a stochastic Lagrangian representation in the 
turbulence modeling field \citep{sawford2001turbulent}. {This stochastic approach is the natural extension to 
diffusive scalars of the Lagrangian description developed for ``ideal'' scalar fields without diffusion advected by smooth velocities.}
We discuss presently this representation only for domains 
$\Omega$ without boundaries, as assumed also by \cite{Bernardetal98}, and in the {companion paper} {II} we describe the extension 
to wall-bounded domains. We shall further discuss in {these papers} only advection by an incompressible fluid satisfying 
\begin{equation} \nabla\cdot \bu=0 \label{div-free} \end{equation}   
so that the ideal advection term formally conserves all integrals of the form $I_h(t)=\int_\Omega d^dx\ h(\theta(\bx,t))$
for any continuous function $h(\theta).$ Note that the representation applies in any space dimension $d,$ with most 
immediate physical interest for $d=2,3,$ of course.


The stochastic representation {of non-ideal scalar dynamics} involves stochastic Lagrangian flow maps $\widetilde{\bxi}_{t,s}^{\nu,\kappa}(\bx)$ 
describing the motion of particles labelled by their positions $\bx$ at time $t$ to random positions at earlier times $s<t.$
The physical relevance of the backward-in-time particle trajectories can be anticipated from the fact that the 
advection-diffusion equation (\ref{eq:PS}) mixes (averages) the values of the scalar field given in the 
past and not, of course, the future values.  Mathematically, the relevant stochastic flows are governed 
by the backward It$\bar{{\rm o}}$ stochastic differential equations:
\begin{equation}\label{noisy}
\hd_s \widetilde{\bxi}_{t,s}^{\nu,\kappa}(\bx) = \bu^\nu(\widetilde{\bxi}_{t,s}^{\nu,\kappa}(\bx),s)\rmd s + 
\sqrt{2\kappa} \ \hd\bW_s, \ \ \ \ \widetilde{\bxi}_{t,t}^{\nu,\kappa}(\bx)= \bx. 
\end{equation} 
Here $\bW_s$ is a standard Brownian motion and $\hd_s$ denotes the backward It$\bar{{\rm o}}$ stochastic differential in the time $s$. 
For detailed discussions of backward It$\bar{{\rm o}}$ equations and stochastic flows, see  \cite{Friedman06,Kunita97}.
For those who are familiar with the more standard forward It$\bar{{\rm o}}$ equations, the backward equations are simply 
the time-reverse of the forward ones. Thus, a backward It$\bar{{\rm o}}$ equation in the time variable $s$ is equivalent to a 
forward It$\bar{{\rm o}}$ equation in the time $\hat{s}=t_r-s$ reflected around a chosen reference time 
$t_r$\footnote{
We note that the difference 
between forward- and backward-It$\bar{{\rm o}}$ equations is not essentially the direction of time in which they are integrated. 
Rather, the difference has to do with the time-direction 
in which those equations are {\it adapted}  \citep{Friedman06,Kunita97}. Thus, a forward It$\bar{{\rm o}}$ differential 
$b(\bW_t)\rmd \bW_t$ is discretized in time as $b(\bW_{t_n})(\bW_{t_{n+1}}-\bW_{t_n})$ for $t_{n+1}>t_n$, with 
the increment $\bW_{t_{n+1}}-\bW_{t_n}$ statistically independent of $\bW_t$ for $t\leq t_n.$ Instead, a backward 
It$\bar{{\rm o}}$ differential $b(\bW_t)\hd \bW_t$ is discretized as $b(\bW_{t_n})(\bW_{t_n}-\bW_{t_{n-1}})$ for $t_n>t_{n-1},$
with $\bW_{t_n}-\bW_{t_{n-1}}$ statistically independent of $\bW_t$ for $t\geq t_n.$ The distinction only matters when,
as in our equation (\ref{Bdiff}), the differential of $\bW_s$ is multiplied by a stochastic function of $\bW.$}. 
The noise term involving the Brownian motion in Eq.(\ref{noisy}) is proportional to the square root of the molecular 
diffusivity $\kappa.$  The velocity field $\bu^\nu$ is assumed to be smooth so long as the parameter $\nu>0.$ 
In the case of greatest physical interest when $\bu^\nu$ is a solution of the incompressible Navier-Stokes equation, then 
$\nu$ represents the kinematic viscosity and we assume, for simplicity of presentation, that there is no blow-up in those solutions.
(See \cite{rezakhanlou2014regular} for weak solutions.) 
Because equation \eqref{noisy}  involves both $\nu$ and $\kappa$, its random solutions $\tbxi_{t,s}^{\nu,\kappa}$ have statistics 
which depend upon those parameters, represented by the superscripts. To avoid a too heavy notation, we omit those superscripts 
and write simply $\tbxi_{t,s}$ unless it is essential to refer to the dependence upon $\nu,\kappa$.  Note that when 
$\kappa=0$ and $\bu^\nu$ remains smooth, then $\bxi_{t,s}^{\nu,0}(\bx)$ is no longer stochastic and gives the usual reverse 
Lagrangian flow from time $t$ backward to the earlier time $s<t$. 

The stochastic representation of the solutions of the advection-diffusion equation follows from the backward differential 
\begin{eqnarray}\label{Bdiff} 
\hat{\rmd}_s \theta(\tbxi_{t,s}(\bx),s) &=& [(\partial_s +\bu^\nu\cdot\nabla-\kappa\Delta)\theta](\tbxi_{t,s}(\bx),s) \rmd s 
+\sqrt{2\kappa} \hd \bW_s\cdot   \nabla \theta(\tbxi_{t,s}(\bx),s) \cr
&=& S(\tbxi_{t,s}(\bx),s) \rmd s +\sqrt{2\kappa} \hd \bW_s\cdot   \nabla \theta(\tbxi_{t,s}(\bx),s), 
\end{eqnarray}
using the backward  It$\bar{{\rm o}}$ formula \citep{Friedman06,Kunita97} in the first line and Eq.(\ref{eq:PS}) in the second. 
Integrating over time $s$ from $0$ to $t,$ gives
\begin{equation}\label{intBdiff}
 \theta(\bx,t)= \theta_0(\tbxi_{t,0}(\bx))+ \int_{0}^t S{(\tbxi_{t,s}(\bx),s) } \ \rmd s 
 + \sqrt{2\kappa} \int_{0}^t \hd \bW_s\cdot   \nabla \theta{(\tbxi_{t,s}(\bx),s) },  
\end{equation}
where $\theta_0$ is the initial data for the scalar at time $0.$  Because the backward It$\bar{{\rm o}}$ integral term 
in (\ref{intBdiff}) averages to zero, one obtains
\begin{equation}\label{Srep}
 \theta(\bx,t)= {\mathbb E}\left[\theta_0(\tbxi_{t,0}(\bx))+ \int_{0}^t S{(\tbxi_{t,s}(\bx),s) } \ \rmd s\right]
\end{equation}
where ${\mathbb E}$ denotes the average over the Brownian motion. Eq.(\ref{Srep}) is the desired stochastic representation
of the solution {of the advection-diffusion equation (\ref{eq:PS})}. Note that the reverse statement 
is also true, that the field $\theta(\bx,t)$ defined {\it a priori} by Eq.(\ref{Srep}) is the solution {of (\ref{eq:PS}) 
for the initial data $\theta_0.$} For a simple proof, see section 4.1 of \cite{EyinkDrivas14} which gives the analogous 
argument for Burgers equation. 

{To see that this stochastic representation naturally generalizes the standard Lagrangian description to 
{non-ideal scalars}, observe that} the scalar values along stochastic Lagrangian trajectories $\theta(\tbxi_{t,s}(\bx),s)$ are,
for $S\equiv 0,$ {\it martingales} backward in time. This means that 
\begin{equation}\label{mart} 
{\mathbb E}\left[\left. \theta(\tbxi_{t,s}(\bx),s)\right| \{\bW_\tau, r<\tau<t\}\right]
= \theta(\tbxi_{t,r}(\bx),r), \quad s<r<t 
\end{equation}
where the expectation is conditioned upon knowledge of the Brownian motion over the time interval $[r,t].$ Thus,
the conditional average value is the last known value (going backward in time).  This is the property for diffusive flow which 
corresponds to the statement for {diffusion-less}, smooth advection that $\theta$ is conserved along Lagrangian trajectories, 
or that $\theta(\bxi_{t,s}(\bx),s)$ is constant in $s$. The proof is obtained by integrating the differential 
(\ref{Bdiff}) over the time-interval $[s,t]$ to obtain   
\begin{equation}
\theta(\tbxi_{t,s}(\bx),s) = \theta(\bx,t)- \sqrt{2\kappa} \int_s^t \hd \bW_\tau\cdot   \nabla \theta{(\tbxi_{t,\tau}(\bx),\tau) }  \ \rmd \tau.
\end{equation}
and then exploiting the corresponding martingale property of the backward It$\bar{{\rm o}}$ integral \citep{Friedman06,Kunita97}. 
It is important to emphasize that the martingale property like (\ref{mart}) does not hold forward in time, which would instead give 
a solution of the negative-diffusion equation with $\kappa$ replaced by $-\kappa<0.$ Thus, the backward-in-time 
martingale property (\ref{mart}) expresses the arrow of time arising from the irreversibility of the diffusion process.  
 
The main result of this paper is a new exact fluctuation-dissipation relation between scalar dissipation 
due to molecular diffusivity and fluctuations associated to stochastic Lagrangian trajectories. To state the 
result, we introduce a stochastic scalar field\footnote{Note that for all $s<t$
the quantity $\tilde{\theta}(\bx,t;s)=\theta(\tbxi_{t,s}(\bx),s)+ \int_{s}^t S{(\tbxi_{t,r}(\bx),r) } \ \rmd r$ is a martingale backward 
in time, by the same argument used above for $S=0$.} 
\begin{equation}\label{tth-def}
\tilde{\theta}(\bx,t)\equiv\theta_0(\tbxi_{t,0}(\bx))+ \int_{0}^t S{(\tbxi_{t,s}(\bx),s)}\ ds
\end{equation}
which, according to Eq.(\ref{Srep}), satisfies $\theta(\bx,t)=\bE[\tilde{\theta}(\bx,t)]$ when averaged over Brownian 
motions. Thus $\tth(\bx,t)$ in (\ref{tth-def}) represents the contribution to $\theta(\bx,t)$ from an individual stochastic 
Lagrangian trajectory as it samples the initial data $\theta_0$ and scalar source $S$ backward in time. Using
this definition and (\ref{Srep}) we can rewrite (\ref{intBdiff}) as 
\begin{equation}\label{intBdiff2}
  \tth(\bx,t) - {\mathbb E}\left[\tth(\bx,t) \right]
  =- \sqrt{2\kappa} \int_{0}^t \hd \bW_s\cdot   \nabla \theta{(\tbxi_{t,s}(\bx),s) }.
  \end{equation}
Squaring this equation and averaging over the Brownian motion gives
\begin{equation}\label{loc-FDR-pre}
  {\rm Var}\left[\tth(\bx,t)\right]
  =   2\kappa \int_{0}^t ds\  \bE\left[ |\nabla \theta{(\tbxi_{t,s}(\bx),s)}|^2\right], 
\end{equation}
where ``Var'' on the lefthand side denotes the stochastic scalar variance in the average over the Brownian motion and on the righthand side 
we have used the It$\bar{{\rm o}}$ isometry (see \cite{oksendal2013stochastic}, section 3.1) to evaluate the mean square of the backward 
It$\bar{{\rm o}}$ integral. If we now average in $\bx$ over the flow domain $\Omega$, use the fact that the stochastic flows 
$\tbxi_{t,s}$ with condition \eqref{div-free} preserve volume, and divide by $1/2$ we obtain 
\begin{equation}\label{FDR}
  \frac{1}{2}\left\langle{\rm Var}\ \tth(t)\right\rangle_\Omega
  =   \kappa \int_{0}^t ds \Big\langle|\nabla \theta{(s)}|^2\Big\rangle_\Omega.
\end{equation}
This is our exact {\it fluctuation-dissipation relation} (FDR). The quantity on the right is just the volume-averaged 
and cumulative (time-integrated) scalar dissipation, and the quantity on the left is (half) the stochastic scalar variance. 
The relation (\ref{FDR}) thus represents a balance between scalar dissipation and the input of scalar fluctuations from the initial scalar field
and the scalar sources, as sampled by stochastic Lagrangian trajectories backward in time.  

It is important to emphasize that the origin of statistical fluctuations in our relation (\ref{FDR}) is {\it not} that 
assumed in most traditional discussions of turbulence, i.e. random ensembles of initial scalar 
fields, of advecting velocity fields, or of stochastic scalar sources. Our FDR (\ref{FDR}) is valid for fixed 
realizations of all of those quantities. {The fluctuating quantity $\tilde{\theta}({\bf x},t)$
which is defined in (\ref{tth-def}) and that appears in our (\ref{FDR}) is an entirely different object from 
the conventional ``turbulent'' scalar fluctuation $\theta'(\bx,t)$. The latter is usually defined by $\theta' := 
 \theta - \langle \theta\rangle,$ where the scalar mean $\langle \theta\rangle$ is taken to be an ensemble- 
 or space/time-average.  Instead, the origin of randomness in $\tilde{\theta}({\bf x},t)$ is the Brownian motion in 
the stochastic flow equation (\ref{noisy}). In special cases, e.g. a dye passively advected by a turbulent flow,
this mathematical Wiener process has direct significance as the description of a physical Brownian motion 
of individual dye molecules in the liquid \citep{Saffman60,Sawfordetal16}. In general, however, the Wiener process 
is simply a means to model the effects of diffusion in a Lagrangian framework. For example, for a temperature 
field there are no ``thermal molecules'' undergoing physical Brownian motion.} 

{Because our FDR is valid for fixed realizations of initial scalar fields, of advecting velocity fields, or of scalar sources,
we are free to average subsequently over random ensembles of these objects.  In this manner we recover from (\ref{FDR}) 
as special cases some known results.} For example, when the scalar source is a random field with zero mean and delta-correlated 
in time, 
\begin{equation} 
\langle \tilde{S}(\bx,t)\tilde{S}(\bx',t')\rangle = 2 C_S(\bx,\bx')\delta(t-t'), \label{delta-source} \end{equation} 
then we recover the steady-state balance equation for the scalar dissipation
\begin{equation}
\langle \kappa |\nabla\theta|^2\rangle_{\Omega,\infty,S}= \frac{1}{V}\int_\Omega d^dx\ C_S(\bx,\bx)
\label{diss-source-bal} \end{equation} 
where the average on the left is over space domain $\Omega,$ an infinite time-interval, and the random source $\tilde{S}.$
This is the standard result usually derived for Gaussian random source fields as an application of the Furutsu-Donsker-Novikov 
theorem (\cite{Frisch95, novikov1965functionals}). 
{We derive it instead as a consequence of a general {\it steady-state FDR}: 
\begin{align}  \label{diss-source-corr-time}
\left\langle \kappa|\nabla \theta|^2\right\rangle_{\Omega,\infty} &= 
\int_{-\infty}^0 \!\!\! \!\!\! dt \ \left\langle \langle \tilde{S}_L(0)\tilde{S}_L(t)\rangle_{\ \mathbb{E},\Omega}^{\! \top}\right\rangle_\infty,
\end{align}
where the random variable $\tilde{S}_L(\bx,s)=S(\tbxi_{0,s}(\bx),s)$ arises by sampling a single 
realization of the source $S$ along stochastic Lagrangian trajectories, $\langle \cdot \rangle_{\ \mathbb{E},\Omega}^{\! \top}$
denotes the truncated correlation function (covariance) in the average over Brownian motion and space domain, and 
$\langle\cdot\rangle_\infty$ an infinite-time average with respect to the release time $0$ of stochastic particles. 
Further averaging 
(\ref{diss-source-corr-time})
over random ensembles of $S$ with delta-covariance (\ref{delta-source}) then gives the steady-state 
balance (\ref{diss-source-bal}). For details, see Appendix  \ref{ensembles}.}
Similar relations hold for freely decaying scalars with no sources but 
random initial scalar fields. For example, when the initial scalar has a uniform random space-gradient, $\tilde{\theta}_0(\bx) 
= \tilde{\mathbf{G}}\cdot \bx$ with isotropic statistics 
\begin{equation} \langle \tilde{\mathbf{G}}\tilde{\mathbf{G}}^\top \rangle_G =G^2 \mathbf{I}, \label{G-isotropy} \end{equation}
then we recover a relation of \cite{Sawfordetal05,Sawfordetal16}
\begin{equation} 
\kappa \int_{0}^t ds \Big\langle|\nabla \theta(s)|^2\Big\rangle_{\Omega,\theta_0}
 =\frac{1}{4}G^2 \ \bE^{1,2} \left\langle \left|\tbxi_{t,0}^{(1)}-\tbxi_{t,0}^{(2)}\right|^2\right\rangle_{\Omega}
\label{sawford-eq} \end{equation}
where the $1,2$ averages are taken over two independent ensembles of Brownian motion. We delay the derivation 
of the {special cases (\ref{diss-source-bal}), (\ref{diss-source-corr-time}), (\ref{sawford-eq})} 
to Appendix \ref{ensembles}, since the proofs require additional material which will be 
introduced in subsequent sections. 

Note, finally, that the result \eqref{loc-FDR-pre} provides a spatially {\it local fluctuation-dissipation relation}, 
which we may write in the form  
\begin{equation}\label{loc-FDR}
  \frac{1}{2t}{\rm Var}\left[\tth(\bx,t)\right]
  =   \left\langle\bE\left[ \kappa|\nabla \theta{(\tbxi_{t,s}(\bx),s)}|^2\right]\right\rangle_t, 
\end{equation}
where on the right {$\langle \cdot\rangle_t$ denotes an average over $s$ in the time interval $[0,t],$}  
carried out along stochastic Lagrangian 
trajectories moving backward-in-time from space-time point $(\bx,t)$. It follows that at short times the local scalar variance
exactly recovers the local scalar dissipation:
\begin{equation}  \lim_{t\rightarrow 0}\frac{1}{2t}{\rm Var}\left[\tth(\bx,t)\right]
  =   \kappa|\nabla \theta(\bx,0)|^2.    \label{loc-FDR-short} \end{equation} 
A substantial spatial correlation between $\frac{1}{2t}{\rm Var}\left[\tth(\bx,t)\right]$ and $\varepsilon_\theta(\bx,t)=\kappa|\nabla\theta(\bx,t)|^2$
should persist for relatively short times $t$. On the other hand, in the long-time limit the local scalar variance 
becomes {\it space-time-independent} and equals
\begin{equation}  \lim_{t\rightarrow\infty}\frac{1}{2t}{\rm Var}\left[\tth(\bx,t)\right]
  =   \left\langle \kappa|\nabla \theta|^2\right\rangle_{\Omega,\infty}
  \quad \mbox{ for all } \bx\in \Omega.  \label{loc-FDR-long} \end{equation}   
To see that Eq. \eqref{loc-FDR-long} should be true, note that the random variables $\tbxi_{t,s}(\bx)\in\Omega$ for each fixed $\bx$ are an ergodic 
random process in the time-variable $s$ for $\kappa>0$. Because of incompressibility of the velocity field and the ergodicity of the 
stochastic Lagrangian flow, the variables $\tbxi_{t,s}(\bx)$ will be nearly uniformly distributed over $\Omega$
at times $s\leq t-\tau$, where $\tau$ is a characteristic scalar mixing time. This time $\tau$ will be at most of the 
order $L^2/\kappa,$ where $L$ is the diameter of the domain, and thus finite for $\kappa>0$, but usually 
much shorter because of advective mixing by the velocity field. For any positive integer $n$
\begin{equation} \lim_{t\rightarrow\infty}\frac{1}{2t}{\rm Var}\left[\tth(\bx,t)\right]
=  \lim_{t\rightarrow\infty}\frac{1}{t}\int_0^{t-n\tau} ds\ \bE\left[ \kappa|\nabla \theta{(\tbxi_{t,s}(\bx),s)}|^2\right],  
\end{equation}
since the corrections are vanishing as $O(n\tau/t).$ By choosing an $n$ sufficiently large but fixed as $t\rightarrow\infty$, 
we can make the righthand side arbitrarily close to 
\begin{equation}  \lim_{t\rightarrow\infty}\frac{1}{t-n\tau}\int_0^{t-n\tau} ds\ \left\langle\kappa|\nabla \theta(\bxi,s)|^2\right\rangle_{\Omega}=
\left\langle \kappa|\nabla \theta(\bxi,s)|^2\right\rangle_{\Omega,\infty}, 
\end{equation}
where the space-time average $\langle\cdot\rangle_{\Omega,\infty}$ on the right is over $\bxi\in \Omega$ and $s\in [0,\infty).$
Since $\lim_{t\rightarrow\infty}\frac{1}{2t}{\rm Var}\left[\tth(\bx,t)\right]$ is independent of the choice of $n,$
we obtain \eqref{loc-FDR-long}. Of course, here we have assumed all of the various infinite-time averages to 
exist, as they shall (at least along subsequences of times $t_k\rightarrow\infty$) if the space-averaged 
scalar dissipation remains a bounded function of time. 

{Note that if the scalar is freely decaying from bounded initial data $\theta_0$, then the variance on the left-hand-side 
of Eq. \eqref{loc-FDR-long} is also bounded. In that case, the long-time-averaged scalar dissipation rate tends to zero, 
which comes as no surprise. In order to have a non-vanishing long-time dissipation, the scalar must be continually supplied 
to the system so that the variance of $\tth(\bx,t)$ grows linearly in time. 
For example, a scalar source $S(\bx,t)$ within the flow domain can provide the necessary scalar input.
In such a case, the variance of $\tilde{\theta}({\bf x},t)$ grows proportionally to time $t$ at long times because 
of the cumulative contribution from the scalar source $S$ in the time-integral $\int_0^t S(\tilde{\bxi}_{t,s},s) \ ds,$
and the long-time average scalar dissipation rate matches the mean input rate of the scalar.  
}  
{In fact, the expression on the righthand side of (\ref{diss-source-corr-time}) arises 
after dividing by $t$ the variance of this time-integral of $S$ and then taking the limit $t\to\infty.$ 
The linear growth of the variance and the expression in (\ref{diss-source-corr-time}) are 
central-limit-theorem results based on the statistical independence of the flow maps 
$\tbxi_{t,s}$ over widely-separated intervals of time $[t,s].$}

\section{Spontaneous Stochasticity of Lagrangian Trajectories}\label{BGK}

We now specialize in this {section} to the source-less case $S\equiv 0,$ in order to make contact with the work 
of \cite{Bernardetal98} on spontaneous stochasticity and anomalous scalar dissipation.
The stochastic representation (\ref{Srep}) simplifies in this case to 
\begin{equation}\label{Srep0}
 \theta(\bx,t)= {\mathbb E}\left[\theta_0(\tbxi_{t,0}^{\nu, \kappa}(\bx))\right]
 =\int d^dx_0 \ \theta_0(\bx_0)\ p^{\nu,\kappa}(\bx_0,0|\bx,t) 
\end{equation}
where we have introduced the backward-in-time transition probability 
\begin{equation}\label{transprob}
p^{\nu,\kappa}(\bx',t'|\bx,t)={\mathbb E}\left[\delta^d(\bx'-\tbxi_{t,t'}^{\nu,\kappa}(\bx))\right] \quad t'<t
\end{equation} 
for the stochastic flow.  As already noted, the stochastic flow preserves volume when the velocity field is divergence-free.
In terms of the transition probability this means that 
\begin{equation} \int p^{\nu,\kappa}(\bx',t'|\bx,t) d^dx =1, \label{vol-cons} \end{equation}
where ${\rm det}(\partial\tbxi_{t,t'}^{\nu,\kappa}(\bx)/\partial\bx)=1$ is used to write $\delta^d(\bx'-\tbxi_{t,t'}^{\nu,\kappa}(\bx))
=\delta^d(\bx-(\tbxi_{t,t'}^{\nu,\kappa})^{-1}(\bx'))$ and perform the integral over $\bx.$ Note that if the limit $\kappa\rightarrow 0$
is taken with $\nu$ fixed (infinite Prandtl number limit), then the stochastic flow \eqref{transprob} becomes deterministic and 
\begin{equation}\label{deltadist} 
p^{\nu,0}(\bx',t'|\bx,t)=\delta^d(\bx'-\bxi_{t,t'}^{\nu,0}(\bx)),  \quad t'<t. 
\end{equation} 
which corresponds to a single deterministic Lagrangian trajectory passing through position $\bx$ at time $t.$

In the Kraichnan model of turbulent advection it was shown by \cite{Bernardetal98} that 
the joint limit $\nu,\kappa\rightarrow 0$ with $Pr=\nu/\kappa$ fixed is non-deterministic and 
corresponds to more than one Lagrangian trajectory passing through space-time point $(\bx,t).$ 
We remind the reader that the Kraichnan model of turbulent advection replaces the Navier-Stokes
solution with a realization drawn from an ensemble of Gaussian random fields $\bu^\nu$ with mean 
zero $\langle \bu^\nu\rangle={\bf 0}$ and covariance satisfying 
\begin{equation} \langle [u_i^\nu(\bx+\br,t)-u_i^\nu(\bx,t)][u_j^\nu(\bx+\br,t')-u_j^\nu(\bx,t')]\rangle 
=D_{ij}^\nu(\br)\delta(t-t') \end{equation}
for a spatial covariance function satisfying $D_{ij}^\nu(\br)=D_{ji}^\nu(\br)$, $ \partial D_{ij}^\nu(\br)/\partial r_j=0,$ and 
\begin{equation}
D_{ii}(\br) \sim \left\{\begin{array}{ll} D_1 r^\xi  & \ell_\nu\ll r\ll L \cr
                                                         D_2 r^2 & r\ll \ell_\nu
                                \end{array} \right. 
\end{equation}                                 
for some $0<\xi<2$, with the effective ``dissipation length''
\begin{equation} \ell_\nu =(D_1/D_2)^{1/(2-\xi)}. \end{equation}      
Note that $D_2\propto \langle |\nabla\bu^\nu|^2\rangle$ and in real turbulence would be proportional to $\varepsilon/\nu,$
where $\varepsilon$ is the viscous energy dissipation. Hence, $D_2\rightarrow \infty$ or $\ell_\nu\rightarrow 0$ with 
$D_1$ fixed is the analogue for the Kraichnan model of the infinite Reynolds-number limit for Navier-Stokes turbulence.
In fact, one can introduce a ``viscosity'' parameter $\nu$ for the Kraichnan model with units of (length)${\,\!}^2$/(time), so 
that $\ell_\nu=(\nu/D_1)^{1/\xi}.$ For any $\nu>0$ the velocity realizations are spatially smooth, but in the limit 
$\nu\rightarrow 0$ they are only H\"older continuous in space with exponent $0<\xi/2<1.$ It is well-known that 
for such ``rough'' limiting velocity fields the solutions of the deterministic initial-value problem 
\begin{equation} d\bxi(s)/ds =\bu(\bxi(s),s), \quad \bxi(t)=\bx \label{limode} \end{equation}
need not be unique and, if not, form a continuum of solutions (e.g. see \cite{Hartman02}). In the Kraichnan model 
it has been proved in the double limit with both $\nu\rightarrow 0$ and $\kappa\rightarrow 0$ that the transition 
probabilities tend to a limiting form 
\begin{equation} p^*(\bx',t'|\bx,t)=\lim_{\nu,\kappa\rightarrow 0} p^{\nu,\kappa}(\bx',t'|\bx,t).  \label{limtran} \end{equation}            
It is important to stress here that no average is taken over $\bu$ in defining these transition probabilities, but only 
an average over Brownian motions in the stochastic flow equations (\ref{noisy}), while the velocity realization is held 
fixed\footnote{It would be less ambiguous to write them as $p^{\nu,\kappa}_\bu(\bx',t'|\bx,t),$ with $\bu$
denoting the fixed flow realization, but this would lead to an even heavier notation.}.
Most importantly, the limiting transition probabilities for the Kraichnan model are {\it not} delta-distributions of the form 
(\ref{deltadist}) but nontrivial probabilities over an ensemble of non-unique solutions of the limiting ODE (\ref{limode})! 
This remarkable phenomenon is called {\it spontaneous stochasticity}. See \cite{Bernardetal98} and the later papers
of \cite{GawedzkiVergassola00,EvandenEijnden00,EvandenEijnden01,Falkovichetal01,LeJanRaimond02,LeJanRaimond04}. 

As shown in those works, spontaneous stochasticity occurs because of the analogue of \cite{Richardson26} dispersion 
in the Kraichnan model, which leads to a loss of influence of the molecular diffusivity $\kappa$ on the separation 
of the perturbed Lagrangian trajectories after a short  time of order $(\kappa^{2-\xi}/D_1)^{1/\xi}.$  
It is important to emphasize that this result does not mean that randomness in the Lagrangian trajectories
suddenly ``appears'' only for $\nu,\kappa=0$ but instead that the randomness {\it persists} even as $\nu,\kappa\rightarrow 0.$
It is thus a phenomenon that can be observed with sequences of positive values, $\nu,\kappa>0,$ for which the velocity field 
is smooth. For the case of a divergence-free velocity that we discuss here, it is furthermore known that the result
does not depend upon the order of limits $\nu\rightarrow 0$ and $\kappa\rightarrow 0$ which can be taken in 
either order or together\footnote{The only delicate case is when $\kappa\rightarrow 0$ first, so that the Prandtl 
number goes to infinity, and then $\nu\rightarrow 0$ subsequently. Since the Brownian motion disappears from 
the stochastic equation (\ref{noisy}) while the velocity field remains smooth, the limiting Lagrangian trajectories
are deterministic. To observe spontaneous stochasticity in that limit one must additionally allow the {\it initial condition}
to be random, e.g. with $\tbxi(t)=\bx+\epsilon \tilde{\brho}$ for a stochastic perturbation $\tilde{\brho}$ drawn from 
some fixed distribution $P(\brho).$ In that case, spontaneous stochasticity appears in the double limit 
with $\epsilon\rightarrow 0$ and $\nu\rightarrow 0$ together and, for a divergence-free velocity $\bu,$ the limiting 
transition probabilities are identical to those obtained for the other limits involving $\kappa\rightarrow 0$. This 
infinite-Prandtl case is discussed carefully by \cite{GawedzkiVergassola00} and \cite{EvandenEijnden00}.}.
See \cite{Bernardetal98,GawedzkiVergassola00,EvandenEijnden00,EvandenEijnden01,Falkovichetal01,LeJanRaimond02,LeJanRaimond04}
for discussions of this point. 

There is empirical evidence for such phenomena also in Navier-Stokes turbulence obtained from 
numerical studies of  2-particle dispersion. \cite{Eyink11} studied stochastic 
Lagrangian particles whose motion is governed by Eq.(\ref{noisy}) in a $1024^3$ DNS at $Re_\lambda=433$ 
and found that the mean-square dispersion becomes independent of $\kappa$ after a short time of order 
$(\kappa/\varepsilon)^{1/2}.$ \cite{bitane2013geometry} studied dispersion of deterministic 
Lagrangian trajectories $(\kappa=0)$ in a $2048^3$ DNS at $Re_\lambda=460$ and a $4096^3$ DNS at $Re_\lambda=730,$ 
and found that the mean-square dispersion becomes independent of the initial separation $r_0$ of particle pairs 
in a short time of order $r_0^{2/3}/\varepsilon^{1/3}.$ The results of these studies provide
evidence of Lagrangian spontaneous stochasticity for Navier-Stokes solutions. {In particular, 
\cite{bitane2013geometry} find consistent Richardson-dispersion statistics for the two Reynolds numbers studied there.} 
{The principal limitation of these previous studies is that they} averaged over the release points $\bx$ of the particles. A particle dispersion 
averaged over release points which remains non-vanishing in the joint limit $\nu,\kappa\rightarrow 0$ is enough 
to infer spontaneous stochasticity for a set of points $\bx$ of non-zero volume measure \citep{Bernardetal98}. However, 
averaging over $\bx$ removes information about the effects of spatial intermittency and the local fluid environment 
on the limiting behavior of the particle distributions $p^{\nu,\kappa}(\bx',t'|\bx,t)$ for specific release locations $\bx.$
{There was some previous study of such spatial intermittency in pair-dispersion by \cite{biferale2005particle,biferale2014intermittency}
but they studied only deterministic Lagrangian particles at small (Kolmogorov-scale) initial separations, and not the stochastic 
Lagrangian particles relevant to our FDR.}

We present here new data obtained from numerical experiments on a high Reynolds-number turbulence simulation in a $2\pi$-periodic box, 
for a couple of representative release points. We use simulation data from the homogeneous, isotropic dataset in the Johns Hopkins 
Turbulence Database (\cite{Li2008,yu2012studying}), publicly available online at 
\url{http://turbulence.pha.jhu.edu}. 
\noindent
It is ideal for our purposes, since the entire time-history of the velocity is stored for a full 
large-scale eddy-turnover time, allowing us to integrate backward in time the flow equations (\ref{noisy}). 
{A significant limitation, however, is that only one Reynolds number is available, $Re\simeq 5058.$}   
One should consider together with the limit $\kappa\rightarrow 0$ also a limit $\nu\rightarrow 0$ so that the Navier-Stokes solution 
$\bu^\nu$ converges to a fixed velocity $\bu$ that is some sort of weak solution of Euler (as always occurs along
a suitable subsequence $\nu_k\rightarrow 0;$ see \cite{lions1996mathematical}, section 4.4)\footnote{{
The necessity of extracting such a subsequence makes the empirical study of spontaneous stochasticity 
quite difficult, as a matter of principle. The compactness argument of \cite{lions1996mathematical} establishes 
existence of subsequences of $\nu_k\to 0$ such that Navier-Stokes solutions $\bu^{\nu_k}(\bx,t)$ converge 
to a fixed limiting velocity field $\bu(\bx,t)$ that is a ``dissipative Euler solution'', in a suitable sense. 
Unfortunately the proof is not constructive and therefore there is currently no concrete computational 
algorithm to generate any specific convergent subsequence.}
Amusingly, the direct experimental observation of spontaneous stochasticity in such a joint limit may be 
easier in quantum mechanics than in turbulent fluids. See \cite{eyink2015quantum}.}. 
{Since no such joint limit $\nu,$ $\kappa\to 0$ can be considered within the given database at one 
fixed value of viscosity, our study of spontaneous stochasticity is based on the assumption that the 
Reynolds numbers is already ``sufficiently large.'' More precisely, we assume that an inertial-range
super-ballistic Richardson-type dispersion of particle pairs released at space-time point $(\bx,t)$ 
will occur at times $|t'-t|>t_c$, with a cross-over time $t_c=\max\{(\varepsilon/\nu)^{1/2},(\varepsilon/\kappa)^{1/2}\},$
and then $p^{\nu,\kappa}(\bx',t'|\bx,t)\simeq p^*(\bx',t'|\bx,t)$ for $|t'-t|\gg t_c.$ Note that for 
$Pr<1$ mean-square dispersion grows diffusively $\propto \kappa |t'-t|$ up to a time 
$t_\kappa=(\varepsilon/\kappa)^{1/2}$ when relative advection begins to dominate at 
the length-scale $\eta_\kappa=(\kappa^3/\varepsilon)^{1/4}$ within the inertial-range. Instead 
for $Pr>1$ particle pairs also separate diffusively initially but then transition to exponential divergence 
$\sim\exp(t/t_\eta) \eta_\kappa,$ with Kolmogorov time $t_\eta=(\varepsilon/\nu)^{1/2},$ 
until the particles separate to the Kolmogorov dissipation scale $\eta=(\nu^3/\epsilon)$ at 
time $\sim t_\eta \log Pr$ when super-ballistic dispersion commences.  The particle 
distributions $p^{\nu,\kappa}(\bx',t'|\bx,t)$ at $|t'-t|\simeq t_c$ will be distinct in these different
cases and will presumably also depend upon the particular values of $\nu,$ $\kappa$ 
even as $\nu,$ $\kappa\to 0.$ However, Richardson dispersion leads to a very rapid 
``forgetting'' of the precise initial data, and thus it is reasonable to expect that 
$p^{\nu,\kappa}(\bx',t'|\bx,t)\simeq p^*(\bx',t'|\bx,t)$ for $|t'-t|\gg t_c.$ With this assumption 
we may study the limiting particle distributions $p^*(\bx',t'|\bx,t)$ in the database at large
but finite Reynolds number. A check on this assumption is provided by the fact that, 
for incompressible flows, the limiting distributions are also expected to be independent 
of Prandtl number $Pr$ \citep{GawedzkiVergassola00,EvandenEijnden00,EvandenEijnden01}. 
By varying $\kappa$ for the fixed $\nu$ in the database, we can change $Pr$ and verify to what  
extent the Prandtl-independence of limiting distributions holds for our numerical results.}

\begin{figure*}
\begin{subfigure}[b]{0.5\linewidth}
    \centering
    \includegraphics[width=1.0\columnwidth]{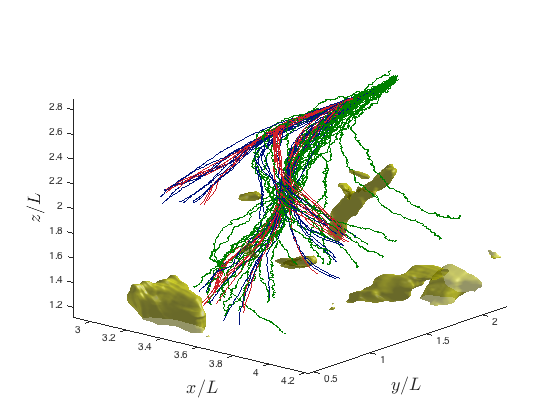} 
   \caption{} 
    \label{fig1:a} 
  \end{subfigure}
  \begin{subfigure}[b]{0.5\linewidth}
    \centering
    \includegraphics[width=1.0\columnwidth]{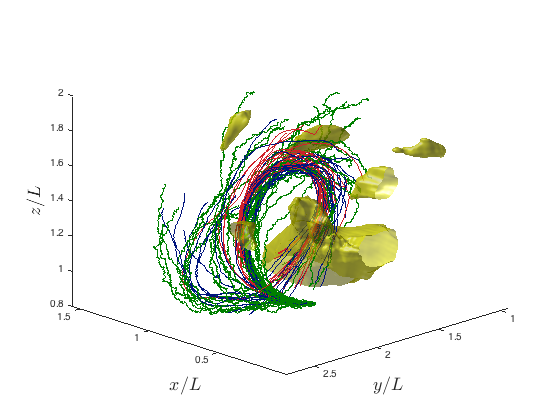} 
   \caption{} 
    \label{fig1:b} 
  \end{subfigure}\\
     \begin{subfigure}[b]{0.5\linewidth}
    \centering
    \includegraphics[width=0.9\columnwidth]{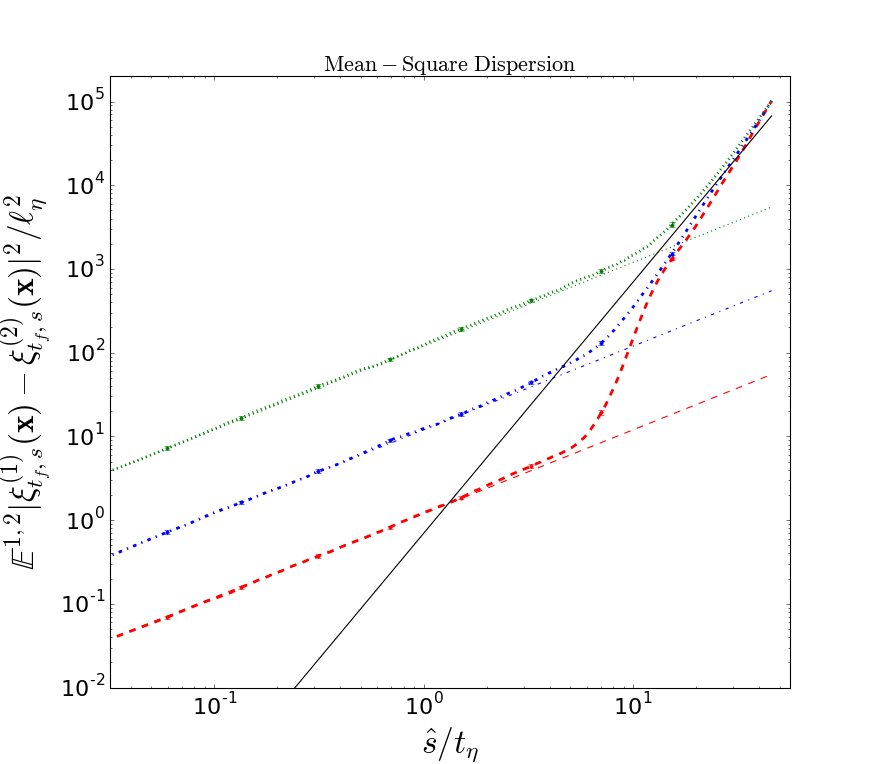} 
   \caption{} 
    \label{fig1:c} 
  \end{subfigure}
  \begin{subfigure}[b]{0.5\linewidth}
    \centering
    \includegraphics[width=0.9\columnwidth]{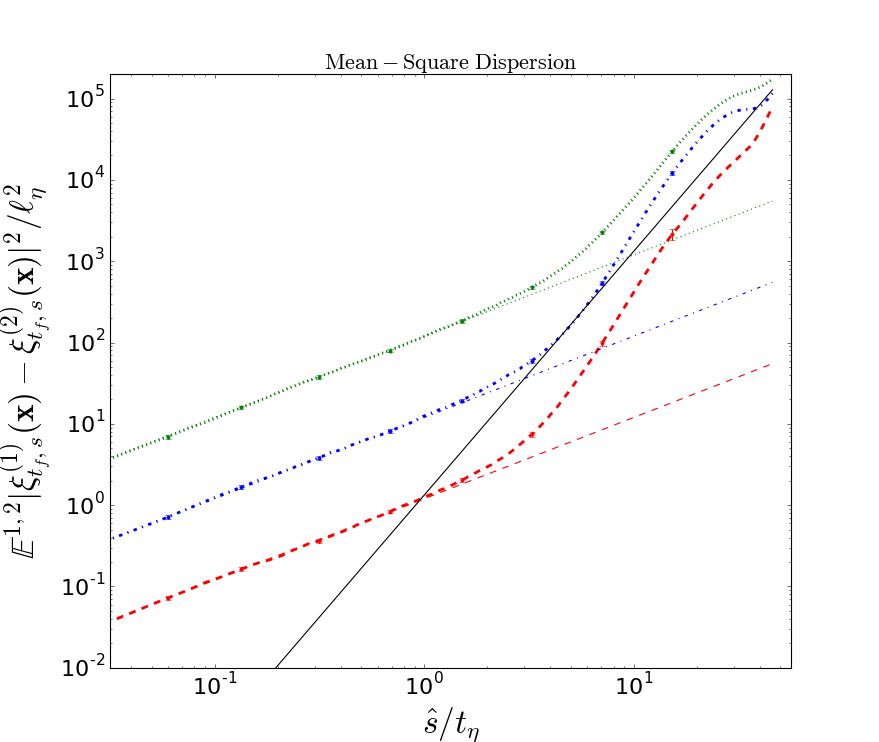} 
   \caption{} 
    \label{fig1:d} 
  \end{subfigure}\\  
  \begin{subfigure}[b]{0.5\linewidth}
    \centering
    \includegraphics[width=0.9\columnwidth]{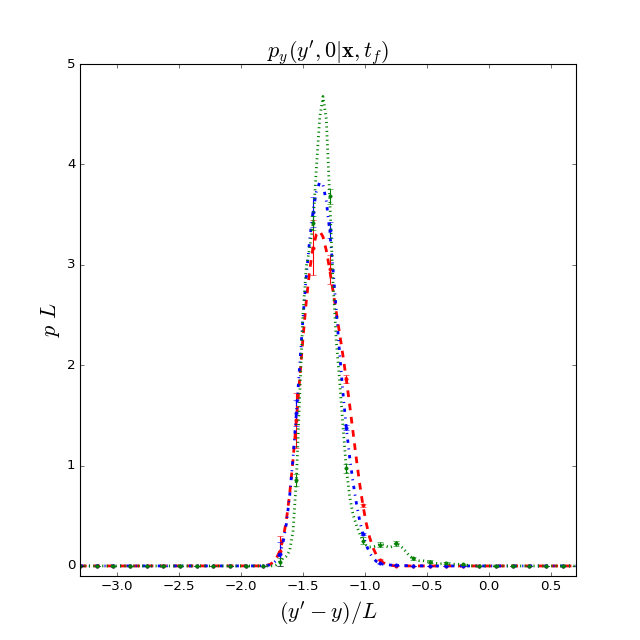} 
   \caption{} 
    \label{fig1:e} 
  \end{subfigure}
  \begin{subfigure}[b]{0.5\linewidth}
    \centering
    \includegraphics[width=0.9\columnwidth]{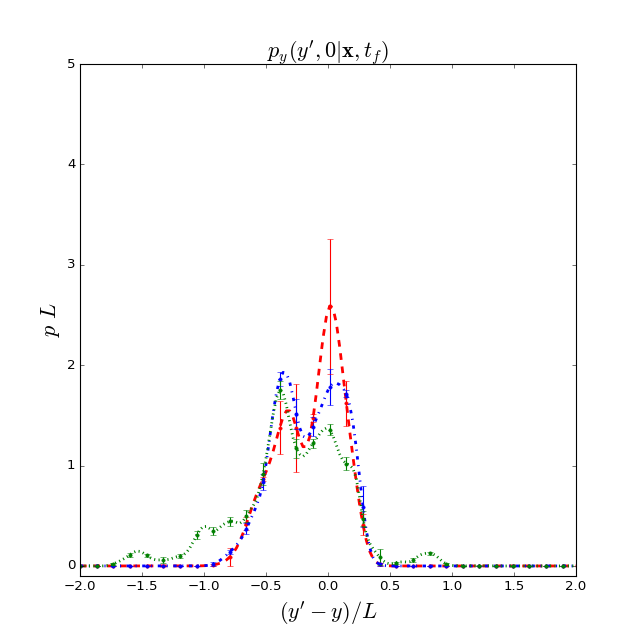} 
   \caption{} 
    \label{fig1:f} 
  \end{subfigure}
\caption{{\it Left panels} are for release at $ \bx=(4.9637, 3.1416, 3.8488)$ in background region, {\it Right panels} for release at 
$\bx=(0.2610, 3.1416, 1.4617)$ near a strong vortex. {Top panels (a),(b) plot 30 representative stochastic trajectories for 
$Pr =0.1$ (\textcolor{ForestGreen}{green}, light), $1.0$ (\textcolor{blue}{blue}, medium) and $10$ (\textcolor{red}{red}, heavy) together with isosurfaces 
of coarse-grained vorticity $|\bar{\bomega}|T_L=15$ at time $s=(2/3) T_{L}.$} Middle panels (c),(d) plot particle 
dispersions (heavy) and short-time results $12\kappa \hat{s}$ (light) for each $Pr$ with $Pr =0.1$ (\textcolor{ForestGreen}{green}, dot, \dottedrule), $1.0$ 
(\textcolor{blue}{blue}, dash-dot, \dasheddottedrule) and $10$ (\textcolor{red}{red}, dash, \dashedrule) and a plot in (solid, \solidrule[4mm]) of $g\varepsilon\hat{s}^3$ 
with $g=0.7$ (left), $g=4/3$ (right). The bottom panels (e),(f) plot $p_y(y',0|\bx,t_f)$ for the three $Pr$-values with the same line-styles as (c),(d).}
\label{fig1}
\end{figure*}

We consider two release points $\bx$ at time $t_f =  2.048,$ the final database time, one chosen in a typical turbulent ``background'' 
region and the other in the vicinity of a strong, large-scale vorticity. We study stochastic trajectories with diffusivities $\kappa$ corresponding 
to three values of the Prandtl number, $Pr =$ 0.1, 1, and 10.  {See Appendix \ref{numerics} for details about the 
numerical methods employed in our analysis.}
In Figure \ref{fig1} the top panels show 30 representative particle trajectories for the two release points and for each of the 
three Prandtl numbers. To illustrate the local fluid environment, we also plot isosurfaces of the vorticity filtered with  
a box-filter of width $L/4$ ($L$ the integral scale) at the time $s=(2/3)T_L$ ($T_L$ the large-scale turnover time). 
The isosurfaces are for magnitudes of filtered vorticity equal to $15/T_L$.
The left panel shows the particles released in a typical ``background'' region with spottier, weaker vortices  
and the right shows particles released near a strong vortex. 
\footnote{{In the weak background region in Fig.~\ref{fig1:a}, $\overline{\omega}_{rms}T_L=2.98$ and thus the isosurface level 
 is $|\overline{\boldsymbol{\omega}}|\approx 5.0\overline{\omega}_{rms}$ with $0.75\%$ of the volume in that box carrying filtered 
vorticity above this threshold. In the strong vortex region in Fig.~\ref{fig1:b} instead $\overline{\omega}_{rms}T_L=4.25$,
so that the isosurface there is $|\overline{\boldsymbol{\omega}}| \approx  3.5\overline{\omega}_{rms}$ with $2.9\%$ of the volume above.}}  
{The three colorings of the trajectories (green/blue/red) represent the three values of the Prandtl numbers $Pr=0.1,1,10$ respectively}. 
{The clearly observable ``splitting'' of the bundle of stochastic trajectories into sub-bundles at specific times recalls 
one proposed mechanism for Richardson dispersion, via a sequence of smooth transport and rapid ``flight-like'' departures
at fluid separatrices \citep{shlesinger1987levy,davila2003richardson,thalabard2014turbulent}.
Most importantly,} as one can see by eye, the ensembles of trajectories are quite similar for the three $Pr$-values. 

To make {the latter} observation more quantitative, we plot in the middle panels of Fig.~\ref{fig1} the mean-square dispersion of 
{pairs of stochastic Lagrangian particles with different realizations of the noise,} for the two release points 
and the three Prandtl numbers. The error bars (almost too small to be observed) represent the standard 
error of the mean (s.e.m.) for averages over N=1024 sample trajectories. For both release points there is an initial period 
(going backward in time) where the dispersion grows diffusively as $12\kappa \hat{s}$ with $\hat{s}=t_f-s,$ but which then crosses over to a 
regime of super-ballistic separation that  is close to the $\hat{s}^3$-growth predicted by \cite{Richardson26} and is approximately 
independent of $Pr$. 
{An essential observation of Fig.~\ref{fig1} is that $\hat{s}\approx t_{c}$ is indeed the time of cross-over 
to a roughly Richardson-$t^3$ growth.}
The two release points shown here illustrate behavior that we have observed also in many other 
points of the turbulent fluid, where we find that the Richardson $\hat{s}^3$-law is {surprisingly} 
robust {(albeit imperfect)}, without the necessity of averaging over release points $\bx.$ This is especially 
so for points $\bx$ in ``background'' regions, and is at least approximately observed for $\bx$ located in more intermittent regions.

Finally, we plot in the bottom panels of Fig.~\ref{fig1} 
particle transition probabilities, which provide even further information about the limiting behavior. We plot 
at time 0, in the approximate Richardson range, the 1-dimensional PDF's of the $y$-coordinate or 
 \begin{equation} p_y^{\nu,\kappa}(y',0|\bx,t_f) = \int dx' dz' \ p^{\nu,\kappa}(\bx',0|\bx,t_f)  \end{equation}  
for each of the two release points $\bx$ and three Prandtl numbers. We observe very similar behavior also for the $x$- and $z$-coordinates. 
In order to minimize the number of samples required to construct the PDF's numerically, we employed kernel density estimator techniques
that gave us good results with only $N=6144$ samples. See \cite{silverman1986density} and Appendix \ref{numerics}, where our 
numerical procedures are completely described. Error bars represent both s.e.m. for the $N$-sample averages and 
the effects of variation in the kernel density bandwidth. Consistent with the dispersion plots, we see that the transition 
PDF's are approximately independent of {$Pr$} for times in the super-ballistic dispersion range.
This is especially true for the release point $\bx$ in the ``background'' region, and for the strong vorticity region such independence 
holds better for the two {largest values of $Pr$ (smallest $\kappa$), when the Richardson-like super-ballistic 
range is the longest. This approximate independence of the PDF's 
from the Prandtl number gives some support to the conjecture that $p^{\nu,\kappa}(\bx',0|\bx,t_f)\simeq p^*(\bx',0|\bx,t_f)$
and that the infinite-$Re$ limit is already achieved for such particle transition kernels in the database at finite $Re$}. 
These numerical studies illustrate the present quality of 
direct evidence for Lagrangian spontaneous stochasticity in high-Reynolds-number Navier-Stokes turbulence,
{which is suggestive but far from compelling.} As we shall
now demonstrate, observations of anomalous scalar dissipation provide further evidence, as the two phenomena are essentially related. 

\section{Spontaneous Stochasticity and Anomalous Dissipation}\label{sec:SS-AD}

{The phenomenon of spontaneous stochasticity leads to a simple explanation 
of anomalous dissipation in a turbulent flow, as was first pointed out by \cite{Bernardetal98} 
for decaying scalars (no sources) in the Kraichnan model of random advection. 
This connection can be {understood} more directly and more generally using our fluctuation-dissipation relation.  
In fact, it is intuitively clear from the FDR \eqref{FDR} that there can be scalar 
dissipation non-vanishing in the limit $\kappa\rightarrow 0$ only if there is a non-vanishing variance 
in that same limit, implying that Lagrangian trajectories must remain 
stochastic. This argument holds in the presence of scalar sources and for a scalar 
advected by any velocity field $\bu^\nu$ whatsoever. In particular, the argument holds when $\bu^\nu$ 
is a Navier-Stokes solution. Thus, spontaneous stochasticity is the only possible mechanism of anomalous dissipation,  
for both passive and active scalars, away from walls. Furthermore, we shall show for a passive scalar which does not react 
back on the flow that spontaneous stochasticity also makes possible anomalous scalar dissipation. Thus, for 
passive scalars the two phenomena are completely equivalent. In this section, we shall deduce
these conclusions, assuming only that the flow domain is compact (closed and bounded) and without any bounding walls.}

We first discuss the technically simpler case with $S\equiv 0$ and then show that the same argument
extends easily to the case with a non-zero scalar source. When $S\equiv 0$ we can rewrite the lefthand side 
of the FDR (\ref{FDR}) using 
\begin{eqnarray}\label{Var2}
&&  {\rm Var}\left[\theta_0(\tbxi_{t,0}(\bx))\right]
   =  \int d^dx_0\int d^dx_0' \ \theta_0(\bx_0) \theta_0(\bx_0') \cr
&&\hspace{35pt}    \times \Big[ p^{\nu,\kappa}_2(\bx_0,0;\bx_0',0|\bx,t)-p^{\nu,\kappa}(\bx_0,0|\bx,t)p^{\nu,\kappa}(\bx_0',0|\bx,t)\Big]. 
\end{eqnarray}
where we have introduced the 2-time (backward-in-time) transition probability density
\begin{equation}
p^{\nu,\kappa}_2(\by,s;\by',s'|\bx,t)=
\bE\left[\delta^d(\by-\tbxi_{t,s}^{\nu,\kappa}(\bx))\delta^d(\by'-\tbxi_{t,s'}^{\nu,\kappa}(\bx))\right], \quad s,s'<t
\end{equation} 
which gives the joint probability for the particle to end up at $\by$ at time $s<t$ and at $\by'$ at time $s'<t,$ 
given that it started at $\bx$ at the final time $t$ (moving backward from final to earlier times). 
At equal times $s=s'$
\begin{equation}
p^{\nu,\kappa}_2(\by,s;\by',s|\bx,t)=\delta^d(\by-\by')p^{\nu,\kappa}(\by,s|\bx,t).
\end{equation}
{We now consider the limit $\nu,\kappa\rightarrow 0$ so that the transition probabilities approach 
limiting values $p^*(\by,s;\by',s|\bx,t)$, $p^*(\by,s|\bx,t).$ Such limits exist, at least along suitably chosen 
subsequences $\nu_n,\kappa_n\rightarrow 0,$ whenever the flow domain is compact. This can be shown 
using Young measure methods similar to those which have been employed previously to study statistical 
equilibria for 2D Euler solutions \citep{robert1991maximum,robert1991statistical,sommeria1991final}. 
Because the proof of this result is a bit technical, we give it in Appendix \ref{Young}. 
When the Lagrangian particles move according to a deterministic flow $\bxi_{t,s}^*$, 
one easily sees that the 2-time transition probability factorizes as  
\begin{equation}
p_2^*(\by,s;\by',s'|\bx,t)=\delta^d(\by-\bxi_{t,s}^*(\bx))\delta^d(\by'-\bxi_{t,s'}^*(\bx))=p^*(\by,s|\bx,t)p^*(\by',s'|\bx,t). 
\end{equation}}
$\!\!\!$Hence, non-factorization in the limit $\nu,\kappa\rightarrow 0$ is an unequivocal sign of spontaneous stochasticity. The variance 
on the lefthand of the FDR (\ref{FDR}) can only be non-vanishing in the limit if factorization fails, so that anomalous dissipation 
clearly requires spontaneous stochasticity. In the other direction, if there is spontaneous stochasticity and thus 
factorization fails for some positive-measure set of $\bx\in \Omega,$ then the contribution to the volume-integrated
variance from that subset must be positive for some suitable smooth choice of $\theta_0,$ which implies a positive lower bound
to the cumulative, volume-integrated scalar dissipation. In short, anomalous scalar dissipation and Lagrangian spontaneous 
stochasticity are seen to be equivalent. This argument is given as a formal mathematical proof in the Appendix 
\ref{proofsSec2}.

{The sufficiency} argument works only for a passive scalar. For active scalars, the initial data $\theta_0$ partially determines the velocity field $\bu$ and so is not free to vary. In order to conclude sufficiency in that case one needs to assume that the resulting velocity field does not ``conspire" with the initial scalar to cause the variance to vanish, i.e. for the random trajectories to sample only points on a single level set of $\theta_0$. If 
this remarkable behavior did happen to occur for some choice of $\theta_0,$ then one would not expect it to persist for a small 
perturbation of $\theta_0.$ Thus, it is highly likely also for active scalars that spontaneous stochasticity implies anomalous dissipation,
but we have not proved that with the FDR. We can however conclude rigorously both for passive and for active scalars that anomalous dissipation implies spontaneous stochasticity. The above proposition shows that any evidence for anomalous scalar dissipation in the free decay of an active or passive scalar (no sources) obtained from DNS in a periodic box is also evidence for spontaneous 
stochasticity. The argument 
in this section is a strong motivation to perform DNS studies to verify anomalous dissipation in the free 
decay of a scalar, since this would provide additional confirmation of spontaneous stochasticity.
All of the DNS cited by \cite{yeung2005high}, section 2.1, employed sources (e.g. a mean scalar gradient 
coupled to the velocity field) that maintained a statistical steady-state for the scalar fluctuations.

Including a non-zero scalar source involves only minor changes to the previous argument. First note that
\begin{eqnarray}\label{Var3}
&&  {\rm Var}\left[\theta_0(\tbxi_{t,0}(\bx))+ \int_{0}^t S{(\tbxi_{t,s}(\bx),s) } \ \rmd s \right]
   ={\rm Var}\left[\theta_0(\tbxi_{t,0}(\bx))\right] \cr 
&& \vspace{10pt}    + 2\ {\rm Cov}\left[\theta_0(\tbxi_{t,0}(\bx)),\int_{0}^t S{(\tbxi_{t,s}(\bx),s) } \ \rmd s \right]
   +{\rm Var}\left[\int_{0}^t S{(\tbxi_{t,s}(\bx),s) } \ \rmd s \right].
\end{eqnarray}
Furthermore, one has for the variance of the time-integrated source sampled along the stochastic particle trajectory that 
\begin{eqnarray}\label{Var4}
&&  {\rm Var}\left[\int_{0}^t S{(\tbxi_{t,s}(\bx),s) } \ \rmd s \right]
   =  \int_{0}^tds\int_{0}^tds' \int d^dy\int d^dy' \ S(\by,s) S(\by',s') \cr
&&\hspace{35pt}    \times \Big[p_2^{\nu,\kappa}(\by,s;\by',s'|\bx,t)-p^{\nu,\kappa}(\by,s|\bx,t)p^{\nu,\kappa}(\by',s'|\bx,t)\Big]. 
\end{eqnarray}
and for the covariance between the sampled initial data and integrated source that 
\begin{eqnarray}\label{Var5}
&&  {\rm Cov}\left[\theta_0(\tbxi_{t,0}(\bx)),\int_{0}^t S{(\tbxi_{t,s}(\bx),s) } \ \rmd s \right]
   =  \int_{0}^tds \int d^dx_0 \int d^dy\ \theta_0(\bx_0)S(\by,s) \cr
&&\hspace{35pt}    \times \Big[p_2^{\nu,\kappa}(\bx_0,0;\by,s|\bx,t)-p^{\nu,\kappa}(\bx_0,0|\bx,t)p^{\nu,\kappa}(\by,s|\bx,t)\Big]. 
\end{eqnarray}
Clearly, anomalous scalar dissipation requires spontaneous stochasticity. For a passive scalar we can also argue 
in the other direction. Indeed, we can repeat the 
previous argument to conclude that, if there is spontaneous stochasticity for a positive measure set of $\bx,$ then
not only is there a smooth choice of $\theta_0$ so that the variance associated to the initial condition in 
(\ref{Var2}) is positive when integrated over this set of $\bx$, but also there is a smooth choice of source field $S$ 
so that the contribution of the variance (\ref{Var4}) is positive. This is already enough to conclude that there must be anomalous 
dissipation for the scalar with initial condition $0$ and with the chosen source $S$. We can also conclude that there 
is  anomalous dissipation for the initial condition $\theta_0$ and the source $S.$ Indeed, if the total variance 
contribution in (\ref{Var3}) is not positive then it must vanish, which implies that the covariance term in 
(\ref{Var5}) provides a negative contribution. In that case, simply take $S\rightarrow -S$ to make the contributions 
of all three terms (\ref{Var2}),(\ref{Var4}),(\ref{Var5}) positive.  We thus conclude that, also for the passive scalar rejuvenated 
by a source,  there is equivalence between anomalous scalar dissipation and Lagrangian spontaneous stochasticity. 
The argument is given more carefully in Appendix \ref{proofsSec2}.

{It has not been generally appreciated that similar conclusions can be reached in the special case of sourceless scalars 
using the arguments of \cite{Bernardetal98}, which are not at all restricted to the Kraichnan model. To underline 
this point and, also, to give additional insight, we here briefly summarize their reasoning.} 
Note that the stochastic representation (\ref{Srep0}) of the advected scalar in the limit $\nu,\kappa\rightarrow 0$ becomes,  
using (\ref{limtran}),  
\begin{equation}\label{Srep00}
 \theta^*(\bx,t) =\int d^dx_0 \ \theta_0(\bx_0)\ p^*(\bx_0,0|\bx,t).  
\end{equation}
{It is worth noting that $\theta^*(\bx,t)$ is a kind of ``weak solution'' of the ideal advection equation, $\partial_t\theta^*+\bu\cdot\nabla\theta^*=0,$
although this fact is not needed for the argument.} 
It follows from \eqref{Srep00} that for any strictly convex function $h(\theta),$ e.g. $h(\theta)=\frac{1}{2}\theta^2,$ 
\begin{equation} h(\theta^*(\bx,t)) \leq  \int d^dx_0 \ h(\theta_0(\bx_0))\ p^*(\bx_0,0|\bx,t), \label{hineq} \end{equation} 
and equality holds if and only the transition probability is a delta-distribution of type (\ref{deltadist}).
This is the so-called Jensen inequality (e.g. see \cite{ito1984introduction}).  Since the limiting transition probabilities are not 
delta-distributions in the Kraichnan model, the inequality in (\ref{hineq}) is strict. Furthermore, the limiting 
transition probabilities for $\nu,\kappa\rightarrow 0$ inherit the volume-preservation property (\ref{vol-cons}), so that 
\begin{equation} \int p^*(\bx',t'|\bx,t) d^dx =1. \label{vol-cons0} \end{equation}  
In that case, integrating (\ref{hineq}) over $\bx$ gives
\begin{equation} \int h(\theta^*(\bx,t)) \ d^dx < \int   h(\theta_0(\bx_0))\ d^dx_0, \label{hineqint} \end{equation} 
so that the $h$-integral is decaying (dissipated) even in the limit $\nu,\kappa\rightarrow 0.$ The anomalous 
scalar dissipation in the Kraichnan model thus has an elegant Lagrangian mechanism. Essentially, the 
molecular diffusivity is replaced by a ``turbulent diffusivity'' associated to the persistent stochasticity 
of the Lagrangian trajectories, which continues to homogenize the scalar field even as the molecular diffusivity 
vanishes. {We give rigorous details of this argument in Appendix \ref{BGKappendix}, where, in the absence of sources, 
we obtain necessary and sufficient conditions for anomalous dissipation identical to those derived from the FDR.}

\section{Summary and Discussion}\label{sec:summary}
 
 This paper has derived a Lagrangian fluctuation-dissipation relation for scalars advected by an incompressible fluid. 
 Our relation expresses an exact balance between  molecular dissipation of scalar fluctuations and the input of scalar fluctuations 
 from the initial scalar values and internal sources as these are sampled by stochastic Lagrangian trajectories 
 backward in time.  We have exploited this relation to give a simple proof (in domains without walls) that spontaneous stochasticity 
 of Lagrangian trajectories is necessary and sufficient for anomalous dissipation of passive scalars, and necessary (but possibly 
 not sufficient) for anomalous dissipation of active scalars. 
 
 {An} important outstanding question is the extent to which the results of this paper can be carried over to provide 
 a Lagrangian picture of anomalous energy dissipation in Navier-Stokes turbulence\footnote{The most direct application of 
our scalar results to Navier-Stokes might appear to be to analyze the viscous dissipation of enstrophy in freely-decaying 2D turbulence,
where the vorticity is an active (pseudo)scalar field. Unfortunately, all  of our analysis assumes that the initial scalar field is 
square-integrable or $L^2,$ but it has been shown by \cite{eyink2001dissipation} and \cite{tran2006vanishing} that there can be no anomalous
enstrophy dissipation for a freely-decaying 2D Navier-Stokes solution with finite initial enstrophy. It may still be the case
that there is anomalous enstrophy dissipation for more singular, infinite-enstrophy initial data and that this dissipation is 
associated to spontaneous stochasticity (see further discussion in \cite{eyink2001dissipation}). However, we cannot investigate this delicate issue 
using the fluctuation-dissipation relation of the present paper.}. We briefly comment upon this issue here. 
  The formal extension of our fluctuation-dissipation relation to viscous energy dissipation is straightforward. We can exploit the stochastic 
 Lagrangian representation for the incompressible Navier-Stokes equation 
 \begin{align}
\partial_t \bu +\bu\cdot \nabla \bu &=-\nabla p+\nu \Delta \bu\label{eq:ns}\\
\nabla \cdot \bu &= 0, \label{continuity}
\end{align}
recently elaborated by \cite{ConstantinIyer08,ConstantinIyer11}, which is valid both for flows in domains without boundaries and for wall-bounded flows. 
Their results can be most simply derived using a backward stochastic particle flow $\tbxi_{t,s} (\bx)$ and a corresponding ``momentum''
$\tbpi_{t,s}(\bx)\equiv  \bu( \tbxi_{t,s}(\bx),s)$ which together satisfy the backward It$\bar{{\rm o}}$ equations
 \begin{align}\label{eq:h1}
\hd \tbxi_{t,s} (\bx) &= \tbpi_{t,s}(\bx)  \rmd s + \sqrt{2\nu} \ \hd \bW_s,\\
\hd \tbpi_{t,s}(\bx) &=-\nabla p(\tbxi_{t,s}(\bx),s)\rmd s+ \sqrt{2\nu} \ \hd  \bW_s\cdot \nabla \bu(\tbxi_{t,s}(\bx),s).\label{eq:h2}
\end{align}
These are a stochastic generalization of Hamilton's particle equations, making contact with traditional methods of Hamiltonian fluid mechanics 
\citep{salmon1988hamiltonian}. See more detailed discussion of \cite{Eyink10,rezakhanlou2014regular}.  By integrating the second of these 
Hamilton's equations from $0$ to $t$ and taking expectations over the Brownian motion, one readily obtains  
\begin{align}
\bu(\bx,t)&= \mathbb{E}\left[\bu_0 (\tbxi_{t,0}(\bx))- \int_{0}^t \nabla p(\tbxi_{t,s}(\bx),s)\rmd s\right]  \label{hamstochrep},
\end{align}  
using the fact that the stochastic integral $\sqrt{2\nu} \int_{0}^t \hd  \bW_s\cdot \nabla \bu^\nu(\tbxi_{t,s}(\bx),s)$ is a backward
martingale and so vanishes under expectation. The formula (\ref{hamstochrep}) was previously derived by \cite{albeverio2010generalized}. 
Moreover, by exploiting the same It$\bar{{\rm o}}$-isometry argument as applied earlier for scalars, one can derive 
 \begin{align}\label{NSVar}
\nu \int_{0}^t \rmd s \  \langle |\nabla \bu(s)|^2\rangle_\Omega  = \frac{1}{2}\left\langle {\rm Var} \left[\bu_0 (\tbxi_{t,0})- 
\int_{0}^t \nabla p(\tbxi_{t,s},s)\rmd s\right]  \right\rangle_\Omega.
\end{align} 
This can be considered a ``fluctuation-dissipation relation'' for viscous energy dissipation in a Navier-Stokes solution. 

Unfortunately this relation does not appear to be particularly useful for analyzing the high-Reynolds number (or inviscid) limit.
It has a mixed Eulerian-Lagrangian character, since it involves both the particle trajectories $\tbxi_{t,s}(\bx)$ 
and the Eulerian pressure-gradient field $\nabla p(\bx,t).$ The latter field is furthermore a dissipation-range object,
which grows increasingly singular as $\nu\rightarrow 0$. For example, using the classical K41 scaling estimates 
\citep{Obukhov49,yaglom49,Batchelor51}, one expects an rms value of the pressure-gradient 
$(\nabla p)_{rms}\sim (\varepsilon^3/\nu)^{1/4}$ and intermittency effects will make this field even more 
singular. Mathematically speaking, the pressure-gradient cannot be expected to exist as an ordinary function in the 
limit $\nu\rightarrow 0$ but only as a distribution. Because of these facts, we cannot derive from (\ref{NSVar})
any relation between anomalous energy dissipation and spontaneous stochasticity for Navier-Stokes turbulence. In particular, 
even if there were anomalous energy dissipation, the limiting stochastic particle trajectories might become deterministic 
as $\nu\rightarrow 0.$  In that case, the variance on the righthand side of (\ref{NSVar}) could remain non-vanishing, 
because the smaller fluctuations due to vanishing stochasticity could be compensated by the diverging magnitude of the pressure-gradient.   

More fundamentally, we believe that (\ref{NSVar}) misses essential physics. Note 
that this relation holds for freely-decaying Navier-Stokes turbulence both in 2D and in 3D, but in the former case
there is certainly no anomalous energy dissipation. Furthermore, in forced, steady-state 2D turbulence there 
is evidence in the inverse-energy cascade range for Richardson dispersion and Lagrangian spontaneous 
stochasticity (\cite{BoffettaSokolov02,FaberVassilicos09}) but this is associated not to small-scale energy dissipation
by viscosity but instead to  large-scale energy dissipation by Eckman-type damping. A possibly important clue is provided 
by the fact that Richardson dispersion is faster backward in time for 3D forward energy cascade \cite{Sawfordetal05,Bergetal06,Eyink11}, 
but faster forward in time for 2D inverse energy cascade \cite{FaberVassilicos09}. By a comparison of these 
observations for 2D and 3D Navier-Stokes turbulence and by means of exact results for Burgers turbulence, 
\cite{EyinkDrivas14} have argued that anomalous energy dissipation for Navier-Stokes turbulence should be related 
not simply to presence of spontaneous stochasticity but instead to time-asymmetry of the stochastic Lagrangian trajectories.  
{This is reminiscent of so-called ``Fluctuation Theorems'' in non-equilibrium statistical mechanics, which 
imply exponential asymmetry in the probability of entropy production with positive and negative signs. 
See \cite{schuster2013nonequilibrium,gawedzki2013fluctuation} for recent reviews. These results are deeply related
to traditional fluctuation-dissipation theorems in statistical physics, but we have been unable to discover 
any connection with our Lagrangian FDR. More recently, a time-asymmetry has been established in the very short-time dispersion 
of nearby Lagrangian trajectories by \cite{falkovich2013single,jucha2014time}.  However, these results hold only for times 
of order {$\sim (r_0^2/\varepsilon)^{1/3}$} and therefore cannot explain the long-time Richardson behavior 
or the observed time-asymmetry therein.}

{The most important implication of the present work is the additional support provided to the concept of Lagrangian
spontaneous stochasticity. Exploiting our Lagrangian FDR, we have shown that any empirical evidence for anomalous 
scalar dissipation, either for passive or for active scalars and away from walls, must be taken as evidence also for spontaneous 
stochasticity. There are profound implications of this phenomenon for many Lagrangian aspects of turbulent flows.
For example, \cite{ConstantinIyer08} have shown that the classical Kelvin-Helmoltz theorems for vorticity dynamics 
in smooth solutions of the incompressible Euler equations generalize within their stochastic framework to solutions 
of the incompressible Navier-Stokes equation with a positive viscosity. In fact, similar to the case of the advected scalars 
discussed in the present work, \cite{ConstantinIyer08} proved that circulations around stochastically advected 
loops are martingales backward in time for the Navier-Stokes solution and also proved that this property completely characterizes 
those solutions.  This ``stochastic Kelvin theorem'' demonstrates again that the stochastic Lagrangian approach is the natural 
generalization to non-ideal fluids of the Lagrangian methods for ideal fluids. Furthermore, if there is spontaneous stochasticity, 
then vortex motion must remain stochastic for arbitrarily high Reynolds numbers. Contrary to the traditional arguments of 
\cite{TaylorGreen37}, vortex-lines in the ideal limit will not be ``frozen-into'' the turbulent fluid flow in the usual sense. 
Similar results holds also for magnetic field-line motion in resistive magnetohydrodynamics \citep{Eyink09}, and spontaneous 
stochasticity then implies the possibility of fast magnetic reconnection in astrophysical plasmas for arbitrarily small electrical conductivity 
\citep{Eyinketal13}. In our companion papers II, III we extend the derivation of our Lagrangian FDR to wall-bounded flows, and 
derive similar relations between anomalous scalar dissipation and spontaneous stochasticity, as well as new Lagrangian
relations for Nusselt-Rayleigh scaling in turbulent convection.}

 
 \section*{Acknowledgements}

We would like to thank Katepalli Sreenivasan for providing us with several important references, as well as Sam Punshon-Smith and Cristian C. Lalescu for useful discussions.  We would also like to thank the Institute for Pure and Applied Mathematics (IPAM) at UCLA, 
where this paper was partially completed during the fall 2014 long program on ``Mathematics of Turbulence''. We also acknowledge
the Johns Hopkins Turbulence Database for the numerical turbulence data employed in this work. G.E. is partially supported by a grant from
NSF CBET-1507469 and T.D. was partially supported by the Duncan Fund and a Fink Award from the Department of Applied Mathematics \& 
Statistics at the Johns Hopkins University.

 \appendix

\section{Mathematical Proofs}\label{proofs}


\subsection{Existence of Limiting Transition Probabilities}\label{Young}

To make rigorous the arguments in Section \ref{sec:SS-AD}, we note that the transition probabilities 
$p^{\nu,\kappa}(\bx_0,0|\bx,t)$ discussed there are well-defined for any sequence of 
continuous (or even just bounded) velocity fields $\bu^\nu.$ However, we shall generally 
assume that these fields are even smooth for $\nu>0$ and their energies are bounded uniformly in $\nu$.  
Because of the latter assumption we can always extract a subsequence $\nu_j\rightarrow 0$ such that $\bu^{\nu_j}\rightarrow \bu,$ 
with $\bu$ a finite energy or $L^2(\Omega\times [0,T])$ velocity field, where convergence is in the weak sense:  
\begin{equation}  \lim_{j\rightarrow\infty}\int_\Omega d^dx\int_0^T dt\ \bu^{\nu_j}(\bx,t)\cdot\bw(\bx,t)=
\int_\Omega d^dx \int_0^T dt\ \bu(\bx,t)\cdot\bw(\bx,t) \end{equation}
for all $\bw\in L^2(\Omega\times [0,T]).$ This is a consequence of the Banach-Alaoglu Theorem \citep{rudin2006functional}. 
Thus, we consider limits in which there is a definite, fixed fluid velocity $\bu.$ If the $\bu^\nu$ 
are solutions of the incompressible Navier-Stokes equation indexed by viscosity $\nu$, then we can 
furthermore select the subsequence $\nu_k\rightarrow 0$ so that the limiting velocity $\bu$ is a ``dissipative Euler solution''
in the sense of \cite{lions1996mathematical}, section 4.4.

We must now show that a further subsequence $\nu_k=\nu_{j_k}$ can be selected together with a corresponding 
subsequence $\kappa_k\rightarrow 0,$ so that the transition probabilities 
$p^{\nu_k,\kappa_k}(\bx_0,0|\bx,t)$ satisfy the following conditions:
\begin{itemize}
\item[(i)] There is a transition density $p^*(\bx_0,0|\bx,t)$ which is measurable in $\bx$ so that 
$$\lim_{k\rightarrow\infty} \int_\Omega d^dx_0\int_\Omega d^dx\ f(\bx_0,\bx) \ p^{\nu_k,\kappa_k}(\bx_0,0|\bx,t)
= \int_\Omega d^dx_0\int_\Omega d^dx\ f(\bx_0,\bx) \ p^*(\bx_0,0|\bx,t), $$
for all continuous functions $f\in C(\Omega\times\Omega).$
\item[(ii)] (normalization) $\int_\Omega d^dx_0\ p^*(\bx_0,0|\bx,t)=1$ for a.e. $\bx\in \Omega.$  
\item[(iii)] (volume-conservation) $\int_\Omega d^dx_0 \int_\Omega d^dx \ g(x_0) \ p^*(\bx_0,0|\bx,t)=\int_\Omega d^dx_0\ g(\bx_0)$
for all continuous $g\in C(\Omega).$  
\end{itemize}
To prove the properties (i)-(iii), the key fact we shall use 
is that the transition probability densities for $\nu,\kappa>0$ can be regarded as {\it Young measures}
\begin{equation}
\mu^{\nu,\kappa,t}_\bx(d\bx_0)=d^dx_0\ p^{\nu,\kappa}(\bx_0,0|\bx,t), 
\label{ymdef} \end{equation}
that is, as probability measures $\mu^{\nu,\kappa,t}_\bx$ on $\Omega$ which are measurably parameterized 
by $\bx\in\Omega.$ Fluid-dynamicists will be familiar with Young measures from theories of long-time 
statistical equilibria for two-dimensional fluids \citep{robert1991maximum,sommeria1991final}. A good 
introduction are the lectures of \cite{valadier1994course} and a comprehensive treatment can be found 
in the monograph of \cite{florescu2012young}. 

Here we briefly review the necessary theory. In the context of our problem, Young measures may be defined as 
families of probability measures $\mu_x$, defined on a 
compact set $Y\subseteq \RR^m$, measurably parametrized by $x\in X\subset\RR^n$, with $X$ also compact.
This uniquely defines a positive Radon measure $\mu$ over $X\times Y$ given on product sets by
\begin{align}
\mu(A\times B) = \int_A \mu_x(B)\  \rmd x.
\end{align}
By construction, $\mu$ satisfies the following identity 
\begin{align}\label{Jirina}
\langle\mu,f\rangle\equiv \int_{X\times Y} f(x,y) \mu(\rmd x,\rmd y) = \int_X\left(\int_Y f(x,y) \  \mu_x(\rmd y) \right) \rmd x,
\end{align}
for any continuous function $f\in C(X\times Y)$. Moreover, for $f\in C(X)$, one has
\begin{equation} \langle\mu,f\rangle = \int_X f(x) \  \rmd x, \end{equation}
that is to say, the projection of $\mu$ on $X$ is $\rmd x,$ the Lebesgue measure. One may alternatively
take these last two properties as the definition of  a Young measure. That is, for any positive Radon measure $\mu$
on $X\times Y$ whose projection on $X$ is $\rmd x$ there is a mapping $x\mapsto \mu_x$ satisfying (\ref{Jirina}). 
This is the content of the so-called Disintegration Theorems \citep{Jirina59,Valadier73}. The mapping
$x\mapsto \mu_x$ is unique Lebesgue almost everywhere. 

Let us denote by ${\mathcal Y}$ the set of Young measures $\mu$ on the product set $X\times Y.$
This set has the important property that it is a closed subset of the space $M(X\times Y)$
of Radon measures on $X\times Y$ in the topology of narrow convergence. The {\it narrow topology}
is the coarsest topology on $M(X\times Y)$ for which the maps $\mu\mapsto \langle\mu,f\rangle$
are continuous for all $f\in C_b(X\times Y),$ the space of bounded continuous functions. 
Since $X\times Y$ is compact, this topology coincides with the so-called {\it vague topology} 
which is the coarsest for which the  maps $\mu\mapsto \langle\mu,f\rangle$
are continuous for all $f\in C_c(X\times Y),$ the space of compactly-supported continuous functions. 
Furthermore, it coincides with the topology defined by the maps $\mu\mapsto\langle\mu,f\rangle$ for all $f\in C(X\times Y).$
For a detailed discussion of these different topologies, see \cite{florescu2012young}. Here we note 
only that these make $M(X\times Y)$ into a compact, metrizable topological space for compact $X,Y.$ 
That ${\mathcal Y}$ is a closed subspace of $M(X\times Y)$ may then easily seen by noting 
that for any sequence $\mu^n\in {\mathcal Y}$ with $\mu^n\rightarrow \mu$ narrowly 
\begin{equation}  \int_X f(x) \  \rmd x =\langle\mu^n,f\rangle \rightarrow \langle\mu,f\rangle, 
\quad \mbox{ for all $f\in C(X)$} \end{equation}
so that the projection of $\mu$ onto $X$ is $\rmd x$ and $\mu\in{\mathcal Y}.$ A further closed subset ${\mathcal Y}_m\subset 
{\mathcal Y}$ is the set of {\it measure-preserving Young measures}, which satisfy the additional 
condition that 
\begin{equation} \langle\mu,g\rangle=\int_X\left(\int_Y g(y) \mu_x(\rmd y) \right) \rmd x =\int_Y g(y) \rmd y, 
\quad \mbox{ for all $g\in C(Y)$ } \end{equation}
which may be stated formally as $\int_X \rmd x \ \mu_x(\rmd y)  =\rmd y.$ That ${\mathcal Y}_m$
is closed in the narrow topology is shown by an argument exactly like that for ${\mathcal Y}$ above. 
  
From these basic results we can easily derive the consequences (i)-(iii), taking $X=Y=\Omega,$
where $\Omega$ is the closure of a bounded open set with a smooth boundary. Then with the 
definition (\ref{ymdef}) one has $\mu^{\nu,\kappa,t}\in {\mathcal Y}_m$ for fixed $t$ and all $\nu,\kappa>0.$
Since ${\mathcal Y}_m$ is a closed subset of the compact, metrizable space $M(X\times Y),$ it
is itself (sequentially) compact. Hence, given the subsequence $\nu_j$ there is a further 
subsequence $\nu_k=\nu_{j_k}$  and a corresponding sequence $\kappa_k$ such that 
$\mu^{\nu_k\kappa_k,t}\rightarrow \mu^{*t}\in {\mathcal Y}_m$ in the narrow topology. Note 
that the limit $\mu^{*t}$ need not be unique and may depend upon the selected subsequence. 
The narrow convergence $\mu^{\nu_k\kappa_k,t}\rightarrow \mu^{*t}$ is equivalent to (i), with the definition
\begin{equation}  d^dx_0\ p^*(\bx_0,0|\bx,t)=\mu^{*,t}_\bx(d\bx_0), \end{equation}
where in general $p^*(\bx_0,0|\bx,t)$ is a distribution in the variable $\bx_0$ not an ordinary function. 
Then (ii) is a restatement that $\mu^{*t}\in {\mathcal Y}$ and (iii) is a restatement that $\mu^{*t}\in 
{\mathcal Y}_m.$ These observations complete the proof of properties (i)-(iii) above.

{
With these results in hand, we now rigorously prove the equivalence of spontaneous stochasticity and anomalous dissipation.  
We do this in two ways: first, by exploiting our general fluctuation dissipation relation and second, by the original argument of 
\cite{Bernardetal98} for the case of scalars without sources.}

\subsection{Proofs Using the FDR}\label{proofsSec2}

As in the main text, we first consider the case without a scalar source ($S=0$). Our starting point is the 
FDR (\ref{FDR}), with formula (\ref{Var2}) for the variance  
${\rm Var}\left[\theta_0(\tbxi_{t,0}^{\nu_k,\kappa_k}(\bx))\right]$. It follows from 
(i)-(ii) of Apppendix \ref{Young} that a subsequence $\nu_k=\nu_{j_k}$ can be selected 
together with a corresponding subsequence $\kappa_k\rightarrow 0,$ so that the space-averaged variance
will satisfy
\begin{eqnarray} \label{limitVar}
&&\lim_{k\rightarrow\infty}  \Big\langle  {\rm Var}\left[\theta_0(\tbxi_{t,0}^{\nu_k,\kappa_k})\right]\Big\rangle_\Omega
   =  \int d^dx \int d^dx_0\int d^dx_0' \ \theta_0(\bx_0) \theta_0(\bx_0') \cr
&&\hspace{35pt}    \times \Big[p_2^{*}(\bx_0,0;\bx_0',0|\bx,t)-p^{*}(\bx_0,0|\bx,t)p^{*}(\bx_0',0|\bx,t)\Big],
\end{eqnarray}
for all $\theta_0\in C(\Omega),$ where 
\begin{equation}  p_2^*(\bx_0,0,\bx_0',0|\bx,t) \equiv \delta^d(\bx_0-\bx_0')p^*(\bx_0,0|\bx,t). \end{equation}
Note that $p_2^*$ is a Young measure on $Y=\Omega\times \Omega$ measurably indexed by elements $\bx$
of $X=\Omega,$ since it is a narrow limit of the Young measures $p_2^{\nu_k,\kappa_k}.$
We shall not use the property (iii) from Appendix \ref{Young} in our argument, although volume-conservation 
was, of course, used in the derivation of the FDR (\ref{FDR}). Since that FDR holds for all $\nu,\kappa>0,$
it follows that the limit of the cumulative global scalar dissipation exists and must coincide with the limiting variance:
\begin{eqnarray} \label{limitEps}
&& \lim_{k\rightarrow\infty}    \kappa_k \int_{0}^t ds \Big\langle|\nabla \theta^{\nu_k,\kappa_k}{(s)}|^2\Big\rangle_\Omega
   =  \int d^dx \int d^dx_0\int d^dx_0' \ \theta_0(\bx_0) \theta_0(\bx_0') \cr
&&\hspace{35pt}    \times \Big[p_2^{*}(\bx_0,0;\bx_0',0|\bx,t)-p^{*}(\bx_0,0|\bx,t)p^{*}(\bx_0',0|\bx,t)\Big], 
\end{eqnarray}
for all $\theta_0\in C(\Omega).$
It follows immediately that anomalous scalar dissipation requires spontaneous stochasticity since, by the exact formula 
\eqref{limitEps}, a non-vanishing cumulative dissipation necessitates non-factorization on a finite measure set of $\bx$. 

The argument that spontaneous stochasticity implies anomalous dissipation is a bit more involved. We need to show that 
if non-factorization holds on a finite measure set of $\bx,$ then there exists a smooth choice of $\theta_0$ such that
both sides of (\ref{limitEps}) are positive. Thus, assume the opposite, that both sides vanish for all smooth 
$\theta_0.$ The righthand size then also vanishes for all continuous $\theta_0,$ since $C^\infty(\Omega)$ is dense 
in $C(\Omega)$ in the uniform norm. For example, this density follows by the Stone-Weierstrass theorem \citep{rudin2006functional}, 
since $C^\infty(\Omega)$ is a subalgebra of $C(\Omega)$ containing the constant $1,$ closed under complex 
conjugation, and separating points of $\Omega.$ Since the integrand with respect to $\bx$ is a variance, it is 
non-negative, so that the vanishing of the integral over $\bx$ implies that there is a subset $\Omega_0\subset\Omega$
of full measure, such that  
\begin{equation}  \int d^dx_0\int d^dx_0' \ \theta_0(\bx_0) \theta_0(\bx_0') 
\Big[p_2^{*}(\bx_0,0;\bx_0',0|\bx,t)-p^{*}(\bx_0,0|\bx,t)p^{*}(\bx_0',0|\bx,t)\Big] =0, \end{equation}
for all $\bx\in \Omega_0$ and $\theta_0\in C(\Omega).$
Note furthermore that the quantity in the square 
brackets ``$[\ \cdot\ ]$'' in the equation above is symmetric in $\bx_0,$ $\bx_0'.$ Thus, 
for any pair of functions $g,h,$ one can take $\theta_0=g+h$ to infer that  
\begin{equation}  \int d^dx_0\int d^dx_0' \ g(\bx_0) h(\bx_0') 
\Big[p_2^{*}(\bx_0,0;\bx_0',0|\bx,t)-p^{*}(\bx_0,0|\bx,t)p^{*}(\bx_0',0|\bx,t)\Big] =0 \end{equation}
for all $\bx\in \Omega_0$ and $g,h\in C(\Omega).$ Since the product functions $(g\otimes h)(\bx,\bx_0')=g(\bx_0)h(\bx_0')$ 
form a subalgebra of $C(\Omega^2)$ that satisfies all of the conditions of the Stone-Weierstrass 
theorem, we can use this theorem again to extend the equality to 
\begin{equation}  \int d^dx_0\int d^dx_0' \ f(\bx_0,\bx_0') 
\Big[p_2^{*}(\bx_0,0;\bx_0',0|\bx,t)-p^{*}(\bx_0,0|\bx,t)p^{*}(\bx_0',0|\bx,t)\Big] =0 \end{equation}
for all $\bx\in \Omega_0$ and $f\in C(\Omega^2).$ The parameterized measure 
$\nu_\bx$ defined by 
\begin{equation}  \nu_\bx(d\bx_0,d\bx_0')=d^dx_0 \ d^dx_0' \Big[p_2^{*}(\bx_0,0;\bx_0',0|\bx,t)-p^{*}(\bx_0,0|\bx,t)p^{*}(\bx_0',0|\bx,t)\Big] \end{equation}
is a difference of two Young measures and, thus, there is a continuous linear functional on $C(\Omega^2)$ 
for all $\bx\in \Omega_0$ also denoted $\nu_\bx,$ defined by $\langle \nu_\bx,f\rangle=\int_{\Omega^2} fd\nu_\bx.$
Since 
\begin{equation}  \langle \nu_\bx,f\rangle =0, \mbox{ for all $f\in C(\Omega^2)$ and $\bx\in \Omega_0,$} \end{equation}
it follows for all $\bx\in\Omega_0$ that  $\nu_\bx\equiv 0,$ as an element of the dual Banach space $C(\Omega^2)^*.$\\
A direct consequence is that 
\begin{equation}  p_2^{*}(\bx_0,0;\bx_0',0|\bx,t)=p^{*}(\bx_0,0|\bx,t)p^{*}(\bx_0',0|\bx,t) \end{equation}
as distributions in $\bx_0,\bx_0',$ for all $\bx\in \Omega_0.$ However, this contradicts our starting assumption that 
factorization fails on a set of full measure. Hence, there must be a smooth choice of $\theta_0$ which makes 
the righthand side of (\ref{limitEps}) positive, and thus also the lefthand side. 

Let us next consider the case with $\theta_0\equiv 0,$ but with the source $S$ non-vanishing. In this circumstance the 
FDR (\ref{FDR}) becomes 
\begin{equation}\label{FDR-S}
\kappa \int_{0}^t ds \Big\langle|\nabla \theta{(s)}|^2\Big\rangle_\Omega
=\frac{1}{2}\left\langle{\rm Var}\left[\int_{0}^t S{(\tbxi_{t,s}^{\nu,\kappa}(s) } \ \rmd s \right]\right\rangle_\Omega
\end{equation}
with expression (\ref{Var4}) for the variance. We show first that there is a suitable subsequence $\nu_k=\nu_{j_k}\rightarrow 0$
and $\kappa_k\rightarrow 0$ such that 
\begin{eqnarray}\label{Var4-lim}
&& \lim_{k\rightarrow\infty}\int_\Omega d^dx  \ {\rm Var}\left[\int_{0}^t S{(\tbxi_{t,s}^{\nu_k,\kappa_k}(\bx),s) } \ \rmd s \right] \cr
&&   \,\,\,\,\,\,\,\,\,\,\,\,\,\,\,\,\,\,\,\,\,\,\,\,\,\,\,\,\,\, 
=  \int_\Omega d^dx\ \int_{0}^tds\int_{0}^tds' \int_\Omega d^dy\int_\Omega d^dy' \ S(\by,s) S(\by',s') \cr
&&\hspace{65pt}    \times \Big[p_2^*(\by,s;\by',s'|\bx,t)-p^*(\by,s|\bx,t)p^*(\by',s'|\bx,t)\Big]. 
\end{eqnarray}
for any $S\in C(\Omega\times [0,t])$ and for suitable limiting transition probabilities $p_2^*$ and $p^*.$ 
To show this we note that 
\begin{equation}  \mu_{s,s',\bx}^{\nu,\kappa}(d\by,d\by') = d^dy\ d^dy' \ p_2^{\nu,\kappa}(\by,s;\by',s'|\bx,t) \end{equation}
defines a set of Young measures on $Y=\Omega\times \Omega$ measurably indexed by elements 
$(s,s',\bx)$ of $X=[0,t]\times [0,t]\times \Omega.$ Since these spaces $X$ and $Y$ are both compact, 
we can appeal to the general results on Young measures discussed in Appendix \ref{Young} to infer that a 
subsequence $\nu_k,\kappa_k$ exists so that, for all $f\in C(X\times Y),$
\begin{eqnarray}
&& \lim_{k\rightarrow\infty} \int_{0}^tds\int_{0}^tds' \int_\Omega d^dy\int_\Omega d^dy' \ \int_\Omega d^dx\ f(\by,s;\by',s';\bx) \
p_2^{\nu_k,\kappa_k}(\by,s;\by',s'|\bx,t) \cr
&& =  \int_{0}^tds\int_{0}^tds' \int_\Omega d^dy\int_\Omega d^dy' \ \int_\Omega d^dx\ f(\by,s;\by',s';\bx) \
p_2^*(\by,s;\by',s'|\bx,t)
\end{eqnarray}
for some limit Young measure with distributional density $p_2^*,$ which it is easy to show inherits the symmetry
of $p_2^{\nu_k,\kappa_k}$ in $(\by,s)$ and $(\by',s').$ Choosing the function $f$ to be of the form 
$f(\by,s;\by',s';\bx)=h(s')g(\by,s;\bx)$ gives also 
\begin{eqnarray}
&& \lim_{k\rightarrow\infty} \int_{0}^tds\int_\Omega d^dy \ \int_\Omega d^dx\ g(\by,s;\bx) \
p^{\nu_k,\kappa_k}(\by,s|\bx,t) \cr
&& \,\,\,\,\,\,\,\,\,\,\,\,\,\,\,\,\,\,\,\,\,\,\,\,\,\,\,\,\,\,\,\,\,\,\,\,\,\,\,\,\,\,\,\,
=  \int_{0}^tds \int_\Omega d^dy \ \int_\Omega d^dx\ g(\by,s;\bx)\ p^*(\by,s|\bx,t) 
\end{eqnarray}
for all continuous $g$ with 
\begin{equation}  p^*(\by,s|\bx,t) = \int_\Omega d^dy'\ p_2^*(\by,s;\by',s'|\bx,t) \end{equation}
constant in $s'$ for almost every $s,\bx$ and defining a consistent 1-time Young measure. We can 
also establish volume-preserving properties of these limiting Young measures, although that will not be 
necessary to our argument. From these results (\ref{Var4-lim}) follows by taking 
the limit along the subsequence $\nu_k,\kappa_k$ of the formula (\ref{Var4}) for the variance. 

The proof that spontaneous stochasticity is both necessary and sufficient for anomalous scalar 
dissipation now follows by arguments almost identical to the situation with $\theta_0\neq 0,$ $S\equiv 0$
that was first considered in this section. Necessity is immediate from (\ref{FDR-S}),(\ref{Var4-lim}). The proof of
sufficiency is very similar to that given before, by showing that vanishing of the space-integrated 
variance (\ref{Var4-lim}) for all smooth source fields $S$ implies the factorization 
\begin{equation}  p_2^*(\by,s;\by',s'|\bx,t)=p^*(\by,s|\bx,t)p^*(\by',s'|\bx,t) \end{equation}
for almost every $\bx\in \Omega.$ The non-negativity of the $\bx$-integrand requires some argument,
because it is no longer obviously a variance. However, it is the limit of a variance in the sense that 
\begin{eqnarray}\label{Var4-lim2}
&& \lim_{k\rightarrow\infty}\int_\Omega d^dx  \ u(\bx) \ {\rm Var}\left[\int_{0}^t S{(\tbxi_{t,s}^{\nu_k,\kappa_k}(\bx),s) } \ \rmd s \right] \cr
&&   \,\,\,\,\,\,\,\,\,\,\,\,\,\,\,\,\,\,\,\,\,\,\,\,\,\,\,\,\,\, 
=  \int_\Omega d^dx\ u(\bx) \ \int_{0}^tds\int_{0}^tds' \int_\Omega d^dy\int_\Omega d^dy' \ S(\by,s) S(\by',s') \cr
&&\hspace{65pt}    \times \Big[p_2^*(\by,s;\by',s'|\bx,t)-p^*(\by,s|\bx,t)p^*(\by',s'|\bx,t)\Big]. 
\end{eqnarray}
for all $u\in C(\Omega)$ and $S\in C(\Omega\times [0,t]).$ If also $u\geq 0,$ then the lefthand side is non-negative
and thus so is the righthand side. This is enough to infer that 
\begin{eqnarray}
&& \int_{0}^tds\int_{0}^tds' \int_\Omega d^dy\int_\Omega d^dy' \ S(\by,s) S(\by',s') \cr 
&&  \,\,\,\,\,\,\,\,\,\,\,\,\,\,\,\,\,\,\,\,\,\,\,\,\,\,\,\, \times 
\Big[p_2^*(\by,s;\by',s'|\bx,t)-p^*(\by,s|\bx,t)p^*(\by',s'|\bx,t)\Big] \geq 0 
\end{eqnarray} 
for all $\bx\in \Omega_0,$ a set of full measure in $\Omega$. The remainder of the argument uses the same 
strategy as before, with $\theta_0\rightarrow S$ and the Banach space $C(\Omega^2)\rightarrow C((\Omega\times [0,t])^2).$

The argument when both $\theta_0\neq 0$ and $S\neq 0$ has already been given in the main text. We only 
add here the technical detail that a single subsequence may be selected so that one has has narrow 
convergence both of the 2-time Young measure 
\begin{equation}  \mu_{s,s',\bx}^{\nu_k,\kappa_k}(d\by,d\by') = d^dy\ d^dy' \ p_2^{\nu_k,\kappa_k}(\by,s;\by',s'|\bx,t)
\rightarrow d^dy\ d^dy' \ p_2^{*}(\by,s;\by',s'|\bx,t)
\end{equation}
and also of the 1-time Young measure at time $t_0=0$
\begin{equation}  \mu_{\bx}^{\nu_k,\kappa_k}(d\bx_0) = d^dx_0 \ p^{\nu_k,\kappa_k}(\bx_0,0|\bx,t)
\rightarrow d^dx_0\ p^{*}(\bx_0,0|\bx,t). \end{equation}
The second statement does {\it not} follow from the narrow convergence 
\begin{equation}  \mu_{s,\bx}^{\nu_k,\kappa_k}(d\by) = d^dy \ p^{\nu_k,\kappa_k}(\by,s|\bx,t)
\rightarrow d^dy\ p^{*}(\by,s|\bx,t)
\end{equation}
because $\{0\}$ is a subset of $[0,t]$ with zero Lebesgue measure. However, after extracting a subsequence 
for which the 2-time Young measure converges, one can extract a further subsequence so that the 
1-time Young measure at time $t_0=0$ also converges.

 \subsection{Rigorous BGK Argument}\label{BGKappendix}
 

We first demonstrate that spontaneous stochasticity implies anomalous dissipation of passive scalars, 
using the same ideas as for the Kraichnan model. We give an indirect proof, supposing that there is no anomalous dissipation and 
showing that there can then be no spontaneous stochasticity, in contradiction to the starting assumption. Thus, we assume 
for some strictly convex function $h$ that 
\begin{equation} \int h(\theta^*(\bx,t)) \ d^dx =\int   h(\theta_0(\bx_0))\ d^dx_0, \end{equation} 
for all smooth initial data $\theta_0.$ Using the volume-conserving property \eqref{vol-cons0},
we can write this equality as 
\begin{equation} \int d^dx \ \left[\int d^dx_0 \  h(\theta_0(\bx_0)) p^*(\bx_0,0|\bx,t) -  h(\theta^*(\bx,t)) \right]=0.  \end{equation}
Because of Jensen's inequality \eqref{hineq} the integrand in the square bracket is non-negative and thus
\begin{equation} h(\theta^*(\bx,t))  = \int d^dx_0 \  h(\theta_0(\bx_0)) p^*(\bx_0,0|\bx,t) \label{loc-heq} \end{equation}
holds pointwise in space for Lebesgue almost every $\bx.$ We can rewrite this equality in terms of the PDF of the random 
variable $\tth(\bx,t)=\theta_0(\tbxi_{t,0}^*(\bx))$ to assume the value $\psi$ or 
\begin{equation}  p_\theta^*(\psi|\bx,t) = \int d^dx_0 \  \delta(\psi-\theta_0(\bx_0)) p^*(\bx_0,0|\bx,t),
\end{equation}
so that \eqref{loc-heq} becomes
\begin{equation} h(\theta^*(\bx,t))  = \int d\psi \  h(\psi) p^*_\theta(\psi|\bx,t) \label{loc-heq-psi} \end{equation} 
with $\theta^*(\bx,t)  = \int d\psi \ \psi\ p^*_\theta(\psi|\bx,t).$ Because of strict convexity of $h,$ Jensen's inequality 
together with \eqref{loc-heq-psi} immediately implies that $p^*_\theta(\psi|\bx,t)$ is a delta-distribution, or 
\begin{equation}  p^*_\theta(\psi|\bx,t) = \delta (\psi-\theta^*(\bx,t)), \label{psi-delta} \end{equation}
and thus $\tth(\bx,t)$ is deterministic. Notice that this conclusion holds for active as well as passive scalars. 
However, \eqref{psi-delta} by itself does not necessarily contradict spontaneous stochasticity, because particle positions $\tbxi_{t,0}^*(\bx)$ 
may remain random but sample only one isosurface of $\theta_0$ for a fixed value $\theta^*$! This latter possibility 
seems very unlikely to be true, even for an active scalar, for every choice of $\theta_0.$ However, we cannot presently 
rule out a possible ``conspiracy'' for an active scalar in which changing $\theta_0$ would always alter 
the velocity field $\bu$ so that the limiting particle positions $\tbxi_{t,0}^*(\bx)$ would remain in an isosurface of $\theta_0$
\footnote{Although our proof does not work for active scalars, we conjecture that if there is some $\theta_0$ for 
which spontaneous stochasticity occurs for an active scalar, then there shall be anomalous scalar dissipation 
for ``generic'' perturbations of $\theta_0$. More formally, in a neighborhood of $\theta_0$ there shall be a dense
$G_\delta$ set of scalar initial data which produce anomalous dissipation. Note that for passive scalars also we 
expect anomalous dissipation for generic $\theta_0,$ although our proofs only guarantee the existence of one 
initial  datum leading to anomalous dissipation.}. For a passive scalar, fortunately, we are free to choose 
$\theta_0(\bx)$ in a completely arbitrary manner without 
altering the velocity field $\bu$ and we can then conclude that the particle positions themselves must be deterministic.
For example, one argument is to take $\theta_0(\bx)=x_i,$ the $i$th-coordinate of $\bx$ in the periodic domain, which 
implies that each $i$th coordinate of $\tbxi_{t,0}^*(\bx)$ must be deterministic\footnote{If the periodic domain has diameter 
$L_i$ in the $i$th direction, we can take $-L_i/2\leq x_i<L_i/2.$ Then $\theta_0(\bx)=x_i$ is clearly not continuous at 
$x_i=\pm L_i/2!$ This technical difficulty can be overcome, if $p^*(\bx_0,0|\bx,t)$ is a measurable function of $\bx_0,$
by choosing sequences of continuous functions $\theta_0(\bx)$ which converge to $x_i$ pointwise. A completely 
different and fully general approach is to integrate \eqref{psi-delta} over $\psi^2,$
or, equivalently, to take $h(\theta)=\theta^2,$ which gives 
$ \int d^dx_0 \ \theta_0^2(\bx_0) \ p^*(\bx_0,0|\bx,t)= \int d^dx_0 \int d^dx_0' \ \theta(\bx_0)\theta(\bx_0') \
p^*(\bx_0,0|\bx,t)p^*(\bx_0,0|\bx,t), $
and to use the argument of the preceding Appendix \ref{proofsSec2}.}. Since this clearly contradicts 
the assumed spontaneous stochasticity of the limiting particle positions, we conclude that for each strictly convex 
function $h,$ there indeed must be anomalous scalar dissipation for some initial data $\theta_0.$

Now assume instead that there is anomalous scalar dissipation, which means that 
the ``deficit'' in the ideally-conserved integral 
\begin{equation} \Delta^{\nu,\kappa}(t) \equiv \int   h(\theta_0(\bx_0))\ d^dx_0 - \int h(\theta^{\nu,\kappa}(\bx,t)) \ d^dx, 
\quad t>0  \label{deficit} \end{equation}
 converges to some limiting value $\Delta(t)>0$ as $\nu,\kappa\rightarrow 0.$ Here we have explicitly indicated
 the dependence of the solution $\theta^{\nu,\kappa}(\bx,t)$ of the scalar advection-diffusion equation upon $\nu,\kappa$.  More precisely, let $\theta_0(\bx_0)$ be continuous on $\Omega$ and consider 
\begin{equation}  \theta^{\nu_k,\kappa_k}(\bx,t)=\int_\Omega d^dx_0\ \theta_0(\bx_0) p^{\nu_k,\kappa_k}(\bx_0,0|\bx,t) \end{equation}
which is measurable in $\bx$ and with scalar energies uniformly bounded as 
\begin{equation}
 \int_\Omega d^dx\ |\theta^{\nu_k,\kappa_k}(\bx,t)|^2 \leq \int_\Omega d^dx_0\ |\theta_0(\bx_0)|^2 
\label{thnukap-ineq} \end{equation} 
by Jensen's inequality and volume-conservation. From (i) {of Appendix \ref{Young}} we see that 
\begin{equation} \lim_{k\rightarrow\infty} \int_\Omega d^dx\ f(\bx) \ \theta^{\nu_k,\kappa_k}(\bx,t)
= \int_\Omega d^dx\ f(\bx) \ \theta^*(\bx,t) \label{lim} \end{equation} 
for all $f\in C(\Omega),$ where we have defined 
\begin{equation} \theta^*(\bx,t)\equiv \int_\Omega d^dx_0\ \theta_0(\bx_0) p^*(\bx_0,0|\bx,t) \end{equation}
which satisfies 
\begin{equation}
\int_\Omega d^dx\ |\theta^*(\bx,t)|^2 \leq \int_\Omega d^dx_0\ |\theta_0(\bx_0)|^2 
\label{thstar-ineq} \end{equation} 
again by Jensen's inequality using (ii) and the volume conservation property (iii) 
{of Appendix \ref{Young}}. Because $C(\Omega)$ is dense in $L^2(\Omega),$
the uniform $L^2$-bounds \eqref{thnukap-ineq},\eqref{thstar-ineq} and the convergence (\ref{lim}) imply that $\theta^{\nu_k,\kappa_k}
\rightarrow\theta^*$ weak in $L^2.$ Integral functionals 
\begin{equation}  H[\theta] = \int_\Omega d^dx \ h(\theta(\bx)) \end{equation}
for convex functions $h$ and finite-measure sets $\Omega$ are weakly lower-semicontinuous on $L^2(\Omega)$ 
(e.g. \cite{berkovitz1974lower}, or \cite{braides2002gamma}, section 2.2), so that one has
\begin{equation}   \int_\Omega d^dx\ h(\theta^*(\bx,t)) \leq \liminf_{k\rightarrow\infty} \int_\Omega d^dx\ h(\theta^{\nu_k,\kappa_k}(\bx,t)). \end{equation}
Now note that if $\Delta^{\nu,\kappa}(t)\rightarrow\Delta(t)$ as $\nu,\kappa\rightarrow 0,$ then 
for all sufficiently small $\nu,\kappa,$ one has $\Delta^{\nu,\kappa}(t)>\Delta(t)/2,$ say, or, in other words, 
 \begin{equation}
 \int_\Omega h(\theta^{\nu,\kappa}(\bx,t)) \ d^dx < \int_\Omega   h(\theta_0(\bx_0))\ d^dx_0 - \frac{1}{2}\Delta(t)
\end{equation} 
Combining this with the above results, we thus obtain
\begin{equation}\label{reverse}
 \int_\Omega h(\theta^*(\bx,t)) \ d^dx \leq \int_\Omega   h(\theta_0(\bx_0))\ d^dx_0 - \frac{1}{2}\Delta(t) < 
 \int_\Omega   h(\theta_0(\bx_0))\ d^dx_0.
\end{equation} 
There must therefore be at least a positive measure set (non-zero volume) of points $\bx$ for which  
\begin{equation} h(\theta^*(\bx,t)) < \int   h(\theta_0(\bx_0))\ p^*(\bx_0,0|\bx,t) \ d^dx_0, \end{equation} 
since otherwise the inequality (\ref{reverse}) would be violated. Because the analogue of (\ref{Srep00})
holds for the limiting transition probability, i.e. 
\begin{equation}\label{Srep00-gen}
 \theta^*(\bx,t) =\int d^dx_0 \ \theta_0(\bx_0)\ p^*(\bx_0,0|\bx,t), 
\end{equation}
and when $h$ is strictly convex, one can conclude that the limiting transition probabilities $p^*(\bx_0,0|\bx,t)$
obtained along the particular subsequence $\nu_n,\kappa_n\rightarrow 0$ are {\it not} delta-distributions 
of type (\ref{deltadist}). Thus, spontaneous stochasticity must hold for at least this positive measure set 
of space points $\bx.$ Note that this direction of the proof did not assume a passive scalar. {\it QED}. 
\

\section{Averages of the Fluctuation-Dissipation Relation over Random Sources and
Initial Data}\label{ensembles}

{
We derive here the specific consequences of our FDR mentioned in section \ref{sect:ss1}.} 

\subsection{{Steady-State Relations \eqref{diss-source-bal} and \eqref{diss-source-corr-time}}}

{We begin with our general steady-state formula Eq. \eqref{diss-source-corr-time}. 
We consider 
compact space domains $\Omega$ without boundary, although it is worth observing 
that identical results hold for wall-bounded domains with no scalar flux through the wall 
{(see Paper II)}. We first note that the contributions from the initial data $\theta_0$ all 
vanish in the limit $t\to\infty,$ so that
\begin{equation}
\left\langle \kappa|\nabla \theta|^2\right\rangle_{\Omega,\infty} 
= \lim_{t\to\infty} \frac{1}{2t} \int_0^t ds \int_0^t ds' \ {\rm Cov}\Big(S(\tbxi_{t,s}(\bx),s),S(\tbxi_{t,s'}(\bx),s')\Big). 
\label{eqB1} \end{equation}
To see this, observe that the finite-time FDR \eqref{FDR} can be expressed as:
\begin{align}\nonumber
 \frac{1}{t} \int_0^t ds \left\langle\kappa|\nabla \theta(s)|^2\right\rangle_{\Omega} \nonumber
   &=\frac{1}{2t} \left\langle{\rm Var}\left[\theta_0(\tbxi_{t,0}(\bx))\right]\right\rangle_{\Omega} \\\nonumber
   &\quad+ \frac{1}{t} \ \left\langle {\rm Cov}\left[\theta_0(\tbxi_{t,0}(\bx)),\int_{0}^t S{(\tbxi_{t,s}(\bx),s) } \ \rmd s \right]\right\rangle_{\Omega} \\
&\quad\quad 
   +\frac{1}{2t} \left\langle  {\rm Var}\left[\int_{0}^t S{(\tbxi_{t,s}(\bx),s) } \ \rmd s \right]\right\rangle_{\Omega}.
\label{eqB2}  \end{align}
For bounded initial data, one has $\frac{1}{2t} {\rm Var}\left[\theta_0(\tbxi_{t,0}(\bx))\right]\leq  {(\max|\theta_0|)^2}/{t} \overset{t\to \infty}{\longrightarrow} 0$.  The contribution of the covariance between the forcing and initial term likewise gives a vanishing contribution, 
since by the Cauchy-Schwartz inequality, 
\begin{eqnarray}
&&  \left| \frac{1}{t}\ {\rm Cov}\left[\theta_0(\tbxi_{t,0}(\bx)),\int_{0}^t S{(\tbxi_{t,s}(\bx),s) } \ \rmd s \right]\right| \cr 
&& \hspace{50pt} \leq 
 \sqrt{\frac{1}{t}{\rm Var}\left[\theta_0(\tbxi_{t,0}(\bx))\right] \cdot\frac{1}{t} {\rm Var}\left[\int_{0}^t S{(\tbxi_{t,s}(\bx),s)} \ \rmd s \right]}.
\end{eqnarray} 
Only the final variance term in (\ref{eqB2}) survives in the limit $t\to\infty.$ Rewriting this using the bilinearity of the 
covariance function gives \eqref{eqB1}. 
}

{Next, using the symmetry in $s,$ $s'$ of the integrand, we can restrict the integration range in \eqref{eqB1} to $s'<s$: 
\begin{equation}
\left\langle \kappa|\nabla \theta|^2\right\rangle_{\Omega,\infty} 
=\lim_{t\to\infty} \frac{1}{t} \int_0^t ds \int_0^s ds' \ {\rm Cov}\Big(S(\tbxi_{t,s}(\bx),s),S(\tbxi_{t,s'}(\bx),s')\Big)
\end{equation}
We then divide the triangular region $R=\{(s,s'): 0<s'<s<t \}$ into three subregions:
\begin{eqnarray}
R_I &=& \{(s,s'): 0<s'<s<t-n\tau \}\cr
R_{II}&=& \{(s,s'): 0<s'<t-2n\tau,\ t-n\tau<s<t \}\cr
R_{III}&=& R\backslash (R_{I}\cup R_{II})  
\end{eqnarray} 
where $\tau$ is the scalar mixing time and $n$ is a positive integer. 
Region $R_{III}$ gives a contribution which is $O(n^2\tau^2/t)$ and can be neglected in the limit $t\rightarrow\infty.$ 
In region $R_{II}$ we can write
\begin{eqnarray}
&& {\rm Cov}\Big(S(\tbxi_{t,s}(\bx),s),S(\tbxi_{t,s'}(\bx),s')\Big) \cr
&&\hspace{25pt} 
=\bE\Big[ \Big( \bE\big(S(\tbxi_{t,s'}(\bx),s')\big| \tbxi_{t,s}(\bx)\big)-\bE\big(S(\tbxi_{t,s'}(\bx),s')\big)\Big) S(\tbxi_{t,s}(\bx),s)\Big], 
\label{eqB6} \end{eqnarray} 
where $\bE\big(\cdot\big| \tbxi_{t,s}(\bx)\big)$ is the conditional average over the Brownian motion given the value of $ \tbxi_{t,s}(\bx)$. 
In region $R_{II}$ both $t-s'>2n\tau$ and $s-s'>n\tau,$ so that for $n\gg 1$ one may use the 
ergodicity of the Lagrangian flow in physical space to obtain 
\begin{equation} 
\bE\big(S(\tbxi_{t,s'}(\bx),s')\big| \tbxi_{t,s}(\bx)\big)\simeq \langle S(s')\rangle_\Omega, 
\quad \bE\big(S(\tbxi_{t,s'}(\bx),s')\big)\simeq \langle S(s')\rangle_\Omega
\end{equation} 
which give nearly canceling contributions in \eqref{eqB6}.  Thus, this region makes an
arbitrarily small contribution for sufficiently large $n.$ Finally, in region $R_I$ we instead write 
\begin{eqnarray}
&& {\rm Cov}\Big(S(\tbxi_{t,s}(\bx),s),S(\tbxi_{t,s'}(\bx),s')\Big) \cr
&&\hspace{15pt} 
=\bE\Big[ S(\tbxi_{t,s}(\bx),s)S(\tbxi_{s,s'}(\tbxi_{t,s}(\bx)),s') \Big]
-\bE\big(S(\tbxi_{t,s}(\bx),s)\big)  \bE\big(S(\tbxi_{t,s'}(\bx),s')\big)  
\label{eqB7} \end{eqnarray} 
using $\tbxi_{t,s'}=\tbxi_{s,s'}\circ \tbxi_{t,s}.$ 
Since $t-s'>t-s>n\tau$ in region $R_I,$ the random variables $\tbxi_{t,s}(\bx),$ $\tbxi_{t,s'}(\bx)$ are 
nearly uniformly distributed over the domain $\Omega$ for $n\gg 1,$ by the  ergodicity of the stochastic 
Lagrangian flow.  By the strong Markov property, one thus gets in region $R_{I}$ that 
\begin{equation}  {\rm Cov}\Big(S(\tbxi_{t,s}(\bx),s),S(\tbxi_{t,s'}(\bx),s')\Big)\simeq 
\langle S(s) \bE\big(S(\tbxi_{s,s'},s')\big)\rangle_{\Omega}
- \langle S(s)\rangle_\Omega\langle S(s')\rangle_\Omega \label{eqB9} \end{equation} 
The righthand side can rewritten more compactly as a ``truncated correlation function'' 
\begin{eqnarray}
&& \langle S(s) \bE\big(S(\tbxi_{s,s'},s')\big)\rangle_{\Omega}
- \langle S(s)\rangle_\Omega\langle S(s')\rangle_\Omega \cr
&&\hspace{25pt} \,= \langle \tilde{S}_L(s,s)\tilde{S}_L(s,s')\rangle_{\bE,\Omega} - 
\langle \tilde{S}_L(s,s)\rangle_{\bE,\Omega}\langle\tilde{S}_L(s,s')\rangle_{\bE,\Omega}  \cr
&& \hspace{25pt} :=\langle \tilde{S}_L(s,s)\tilde{S}_L(s,s')\rangle_{\bE,\Omega}^\top, 
\end{eqnarray} 
where we have defined the Lagrangian source field $\tilde{S}_L(\bx,s,s')=S(\tbxi_{s,s'}(\bx),s')$ as sampled along stochastic 
trajectories for $s'<s$ and we have also introduced the notation $\langle \cdot\rangle_{\bE,\Omega}$ for the joint 
average over Brownian motion and space domain. When the difference of the two sides of \eqref{eqB9} is 
a function integrable over infinite ranges and vanishing as $n\to\infty,$ dominated convergence gives   
\begin{equation}
\left\langle \kappa|\nabla \theta|^2\right\rangle_{\Omega,\infty} 
= \lim_{n\to\infty}\lim_{t\to\infty} \frac{1}{t} \int_0^{t-n\tau} ds \int_0^s ds' \ \langle \tilde{S}_L(s,s)\tilde{S}_L(s,s')\rangle_{\bE,\Omega}^\top 
\end{equation} 
This non-vanishing contribution from region $R_I$ can be combined with vanishing contributions 
from regions $R_{II},$ $R_{III}$ by the reverse of the preceding argument, to give 
\begin{equation}
\left\langle \kappa|\nabla \theta|^2\right\rangle_{\Omega,\infty} 
= \lim_{t\to\infty} \frac{1}{t} \int_0^t ds \int_0^s ds' \ \langle \tilde{S}_L(s,s)\tilde{S}_L(s,s')\rangle_{\bE,\Omega}^\top
\end{equation}
}

{To obtain the final result, we make the change of variables $s'\to \sigma=s'-s,$ giving
\begin{eqnarray}
\left\langle \kappa|\nabla \theta|^2\right\rangle_{\Omega,\infty} 
&=& \lim_{t\to\infty} \frac{1}{t} \int_0^t ds \int_{-s}^0 d\sigma \ \langle \tilde{S}_L(s,s)\tilde{S}_L(s,s+\sigma)\rangle_{\bE,\Omega}^\top \cr 
&=& \lim_{t\to\infty} \frac{1}{t} \int_{-t}^0 d\sigma \int_{-\sigma}^t ds \ \langle \tilde{S}_L(s,s)\tilde{S}_L(s,s+\sigma)\rangle_{\bE,\Omega}^\top 
\label{eqB13}
\end{eqnarray}
after switching the order of integrations. This can be rewritten as 
\begin{equation}
\left\langle \kappa|\nabla \theta|^2\right\rangle_{\Omega,\infty} 
=\lim_{t\to\infty} \int_{-t}^0 d\sigma \ \left(1+\frac{\sigma}{t}\right)
\left\langle \langle \tilde{S}_L(0,0)\tilde{S}_L(0,\sigma)\rangle_{\bE,\Omega}^\top \right\rangle_{[-\sigma,t]}
\label{eqB14} \end{equation} 
where we have introduced the time-average over the interval $[-\sigma,t],$ $\sigma<0:$ 
\begin{equation}
\left\langle f(0) \right\rangle_{[-\sigma,t]}:= \frac{1}{t+\sigma}\int_{-\sigma}^t ds \ f(s)
\end{equation} 
Assuming that the integrand in \eqref{eqB14} is absolutely integrable uniformly in $t$, then
dominated convergence theorem applies and we obtain:
\begin{align} 
&\left\langle \kappa|\nabla \theta|^2\right\rangle_{\Omega,\infty}  =\int_{-\infty}^0\!\!\!\! d\sigma \ 
\left\langle\langle \tilde{S}_L(0,0) \tilde{S}_L(0,\sigma) \rangle_{\bE,\Omega}^\top
\right\rangle_{\infty}. \label{longtimelimeFormula}
\end{align}
Expression \eqref{longtimelimeFormula} is equivalent to formula \eqref{diss-source-corr-time} in the main text.}  \\

{
We now derive as a special case of \eqref{longtimelimeFormula} the result \eqref{diss-source-bal} for 
a statistical steady-state maintained by a random scalar source, delta-correlated in time. 
In addition to the delta-in-time covariance \eqref{delta-source}, the source is assumed to satisfy 
the condition that the ensemble-average is zero and also that the space-integral is zero in every realization: 
\begin{equation}     \int_\Omega d^dx \ \tilde{S}(\bx,t) = 0 \quad a.s. \label{no-input} \end{equation} 
This latter condition means that there is no net input of scalar by the source.  As a consequence, the truncation terms in \eqref{longtimelimeFormula} vanish identically.  Averaging the formula \eqref{longtimelimeFormula} over the random 
source, the delta-covariance \eqref{delta-source} then implies that 
 \begin{equation} \kappa \Big\langle|\nabla \theta|^2\Big\rangle_{\Omega,\infty,S} = \frac{1}{V}\int_\Omega d^dx\ C_S(\bx,\bx).  
 \end{equation}
where integration of the delta-function $\delta(\sigma)$ over $\sigma\in [-\infty,0]$ gives a factor of $1/2.$}
  
\subsection{{Free Decay Relation \eqref{sawford-eq}}}  
  
Next we derive from our general FDR the relation \eqref{sawford-eq} of \cite{Sawfordetal05,Sawfordetal16} for a decaying passive scalar 
with a random initial linear profile satisfying the statistical isotropy condition \eqref{G-isotropy}. To obtain this result,  note that for 
any random variable $\tilde{X}$, ${\rm Var}(\tilde{X}) =\frac{1}{2}\bE\  |\tilde{X}^{(1)}-\tilde{X}^{(2)}|^2$ where $\tilde{X}^{(1)},
\tilde{X}^{(2)}$ are two independent random variables identically distributed as $\tilde{X}$.  Therefore, our FDR \eqref{FDR} 
in the case of vanishing scalar sources and random scalar initial-values can be re-expressed as:
\begin{align}\label{LSF-FDR}
\kappa \int_{0}^t ds\ \Big\langle|\nabla \tth(s)|^2\Big\rangle_{\Omega}
  &   =\frac{1}{4}\left\langle \bE\left|\tth_0(\tbxi_{t,0}^{(1)})- \tth_0(\tbxi_{t,0}^{(2)})\right|^2\right\rangle_{\Omega}.
\end{align}
Assuming that $\tth_0(\bx)=\tilde{{\bf G}}\cdot \bx$ and averaging over random initial data using  \eqref{G-isotropy} gives 
\begin{equation} 
\kappa \int_{0}^t ds \Big\langle|\nabla \tth(s)|^2\Big\rangle_{\Omega,\theta_0}
 =\frac{1}{4}G^2 \  \left\langle \bE^{1,2}\left|\tbxi_{t,0}^{(1)}-\tbxi_{t,0}^{(2)}\right|^2\right\rangle_{\Omega},
\label{sawford-app}  \end{equation}
which is \eqref{sawford-eq}.  

The interest of this relation is that it directly connects the temporal evolution of the mean scalar 
dissipation to two-particle dispersion of stochastic Lagrangian trajectories. For example, it relates dissipative anomalies of scalar 
fluctuations and kinetic energy if the Prandtl number is fixed and if the dispersion on the right of \eqref{sawford-app}
exhibits a Richardson scaling $\sim \varepsilon t^3$ with $\varepsilon$ independent of $\nu$.  This relation can be used 
to establish equivalence of spontaneous stochasticity and anomalous scalar dissipation for situations that satisfy the specific assumptions
under which it is derived.  One can obtain such a relation for more general initial data than linear profiles by instead assuming only  
that the initial scalar field is smooth and that its 2nd-order structure function satisfies the pair of inequalities for all $\bx,\bx'\in \Omega$ that
\begin{equation}
c_{\theta_0}|\bx-\bx'|^2\leq \big\langle |\tth_0(\bx)-\tth_0(\bx')|^2\big\rangle_{\theta_0} \leq C_{\theta_0}|\bx-\bx'|^2, 
\label{scalstrucfun} \end{equation} 
for some constants $0<c_{\theta_0}<C_{\theta_0}<\infty.$ Averaging \eqref{LSF-FDR} over such initial data then yields
\begin{align}
\frac{1}{4}c_{\theta_0}\left\langle \bE\left|\tbxi_{t,0}^{(1)}- \tbxi_{t,0}^{(2)}\right|^2\right\rangle_{\Omega}  
\leq \kappa \int_{0}^t ds\ \Big\langle|\nabla \tth(s)|^2\Big\rangle_{\Omega,\theta_0}   
\leq \frac{1}{4}C_{\theta_0}\left\langle \bE\left|\tbxi_{t,0}^{(1)}- \tbxi_{t,0}^{(2)}\right|^2\right\rangle_{\Omega}.
\end{align}
This gives upper and lower bounds for the cumulative scalar dissipation directly in terms of two-particle dispersion,
which again relate anomalous scalar dissipation to spontaneous stochasticity.  If there is a smooth, random scalar source 
whose structure-function satisfies bounds similar to \eqref{scalstrucfun}, then one can obtain analogous bounds relating 
cumulative scalar dissipation to the time-integrated 2-particle dispersion. 


\section{Numerical Methods}\label{numerics}

\subsection{Methods for Section \ref{BGK} }\label{BGKnumerics}

In order to integrate backward It$\bar{{\rm o}}$ SDE's \eqref{noisy} of the form 
 \begin{equation}\label{noisyAppend}
  \hd\tbxi(s) = \bu(\tbxi(s),s)ds + \sqrt{2\kappa} \ \hd\bW(s) 
  \end{equation}
we use a reflected time $\hat{s}=t_f-s$ which converts them into forward It$\bar{{\rm o}}$ SDE's. The latter 
are integrated with the standard Euler-Maruyama scheme \citep{pe1992numerical}. 
 {We solved Eq.(\ref{noisyAppend}) with the Euler-Maruyama 
scheme, which for additive noise is 1st-order in both weak and strong sense \citep{pe1992numerical}. The turbulent velocity field 
in Eq.(\ref{noisyAppend}) was retrieved from the Johns Hopkins turbulence database with the {\tt getVelocity} function, which returns 
velocities at requested points interpolated in space by 6th-order Lagrange polynomials and in time by piecewise-cubic Hermite polynomials. 
We used time-step $\Delta s = 6.6\times10^{-4},$ or 1/3 of the time between database frames. 
We calculated statistics with averages over independent solutions of Eq.(\ref{noisyAppend}) and,
to test for weak convergence in the time-integration, we doubled $\Delta s,$ with relative change $<0.1\%$.}

{To estimate particle dispersions and transition probability densities 
$p_y(y',0|\bx,t_f)$ we used} $N$-sample ensembles of stochastic trajectories $\tbxi_n(s),$ $n=1,..,N$ solving the above SDE. 
The particle dispersions were calculated by the unbiased estimators 
\begin{eqnarray}
\bE^{1,2}\left[ |\tbxi^{(1)}(s)-\tbxi^{(2)}(s)|^2\right] &\doteq& \frac{2}{N(N-1)}\sum_{n<m} |\tbxi_n(s)-\tbxi_m(s)|^2\cr 
&=& \frac{2}{N-1} \sum_{n=1}^N |\tbxi_n(s)-\overline{\bxi}_N(s)|^2
\end{eqnarray}
with $\overline{\bxi}_N(s)=\frac{1}{N}\sum_{n=1}^N \tbxi_n(s)$ the sample mean. For large $N$ these are nearly the same  as
 \begin{equation} \bE^{1,2}\left[ |\tbxi^{(1)}(s)-\tbxi^{(2)}(s)|^2\right] \doteq \frac{2}{N} \sum_{n=1}^N |\tbxi_n(s)-\bE[\tbxi(s)]]|^2  \end{equation}
which is twice the sample average over $N$ independent random variables each with the same distribution as $|\tbxi(s)-\bE[\tbxi(s)]]|^2$. 
Our error bars for the particle dispersion are thus taken to be twice the standard error of the mean (s.e.m.) for this 
$N$-sample average. 
 
\begin{figure}
\centering    
	\includegraphics[width=.9\linewidth,height=.4\linewidth]{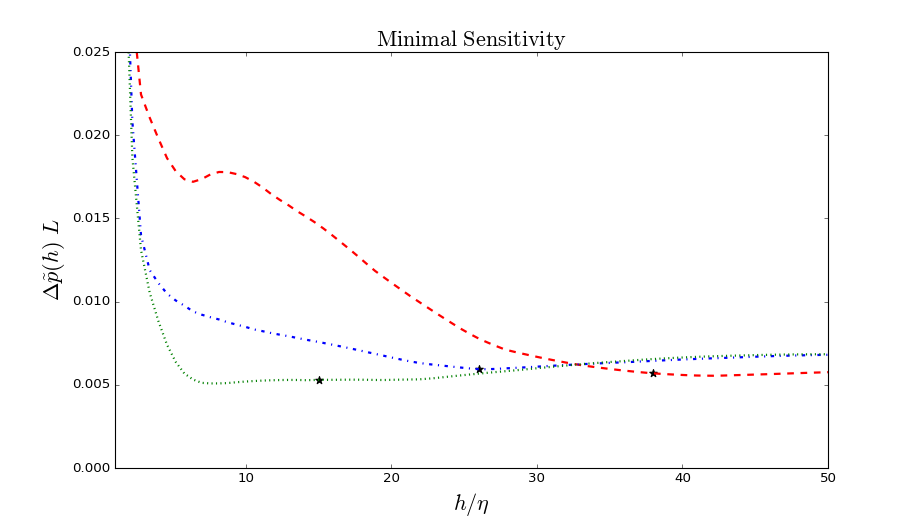}
	\caption{$L^1$-differences of PDF estimates for successive bandwidths $h_j$ are plotted 
	for $Pr = 0.1$ (\textcolor{ForestGreen}{green}, dot, \dottedrule), $1.0$ (\textcolor{blue}{blue}, dash-dot, \dasheddottedrule) and $10$ (\textcolor{red}{red}, dash, \dashedrule). 
	 The optimum bandwidths correspond to the local minima at smallest $h_j$-values, marked with a
	 filled star ($\star$) on the plots.}
\label{fig:minimalSens}
\end{figure}

To estimate the position PDF's, we used kernel density estimator (KDE) methods \citep{silverman1986density}. 
For the $y$-coordinate at time $0$ of the particle $\tbxi(0)=(\tilde{\xi}(0),\tilde{\eta}(0),\tilde{\zeta}(0))$ started at $\bx$ 
at time $t_f,$ one has for a grid $y_i'$ of possible $y$-values
\begin{eqnarray}
\tilde{p}_h(y'_i,0|\bx,t_f) \doteq \frac{1}{N}\sum_{n=1}^N K_h (\tilde{\eta}_n(0) - y_i')
\label{kde} \end{eqnarray}
where $K_h(y)=(1/h)K(y/h)$ is a filter kernel with bandwidth $h.$ We take $K$ to be a Gaussian with unit variance and 
we choose the bandwidth $h$ by the ``principle of minimal sensitivity" from renormalization-group theory \citep{stevenson1981optimized}. 
The latter procedure is based upon the observation that, when the number $N$ of samples is sufficiently 
large for the average in \eqref{kde} to be converged to the convolution $ (K_h*p)(y_i',0|\bx,t_f),$ then the result will
be independent of $h$ for any value less than the scale of variation of the limit PDF. Since this is an exact 
invariance property of the limiting result, the ``principle of minimal sensitivity'' selects the optimal bandwidth $h_*$ 
for finite $N$ so that varying the bandwidth has minimal effect on the PDF-estimate.  Precisely, one picks $h_*$ 
by considering a decreasing sequence of candidate values $h_j,$ computing the $L_1$-difference 
$\Delta\tilde{p}(h_j)  := \|\tilde{p}_{h_j} - \tilde{p}_{h_{j-1}}\|_{L^1}$ for successive bandwidths, and picking $h_*$ where 
$\Delta\tilde{p}(h_j)$ is most nearly flat.  This procedure is illustrated in Figure \ref{fig:minimalSens} for the 
particle position PDF's that were presented in Fig.~\ref{fig1}(f) but using $N=1024$ samples. 
The $L^1$-differences are plotted versus $h_j$ in Figure \ref{fig:minimalSens} for the three choices of Prandtl number.
The bandwidths chosen correspond to the local minima for each curve at the smallest $h_j$-value indicated 
by the star $(\star)$ on the graph, i.e. $h_*/\eta \approx 16$ 
for $Pr =0.1$, $h_*/\eta\approx 26$ for $Pr =1$, $h_*/\eta \approx 39$ for $Pr =10$. In some cases
we did not observe local minima as in Fig.~\ref{fig:minimalSens} and in those instances our procedure 
was to select as ``optimal'' bandwidth $h_*$ the smallest $h_j$ in an interval where the plot of $\Delta\tilde{p}(h_j)$
vs. $h_j$ had a slope of magnitude less than 0.01. 

Finally, after selecting the optimal bandthwidth $h_*,$ we obtained the sample-size error for the PDF estimate 
as the standard error of the mean for the sample average (\ref{kde}) over $N$ independent random variables 
identically distributed as $K_{h_*} (\tilde{\eta}(0) - y_i')$, with bandwidth $h_*$ fixed. In addition to this statistical 
error, an additional source of error arises from the choice of $h_*.$ To assess this, we recalculated the kernel density 
estimator with a 10\% increase in bandwidth, or $1.1h_*,$ and took the absolute difference in the two PDF estimates as a measure 
of the error associated to small variations in band-width. The two types of errors were found to have more or less comparable 
magnitudes and the error bars in Fig.~\ref{fig1}(e),(f) represent the total error obtained from their sum.  

\bibliographystyle{jfm}

\bibliography{bibliography}

\end{document}